\documentclass[onecolumn,floatfix,aps,superscriptaddress]{revtex4}
\usepackage{graphicx}
\usepackage{amsmath}
\usepackage[colorlinks=true, citecolor=blue, urlcolor=blue, linkcolor = blue]{hyperref}
\usepackage{amssymb}
\usepackage{bm}
\usepackage{color}
\usepackage{bookmark}
\usepackage{tabularx}
\usepackage{mathtools}
\usepackage{microtype}
\usepackage{cancel}
\usepackage{relsize}

\usepackage{xcolor}

\begin{document}
\title{Linear and nonlinear characteristics of high frequency electrostatic drift waves in absence of collisions}\author{S. P. Acharya
\affiliation{Saha Institute of Nuclear Physics, 1/AF Bidhannagar, Kolkata-700064 (India)}
\footnote{Email: sibaprasad.acharya@ipr.res.in and siba.acharya39@gmail.com}\footnote{Current affiliation: Institute for Plasma Research, Bhat, Gandhinagar-382428, Gujarat (India)}}

\author{M. S. Janaki
\footnote{Email: ms.janaki.retd@saha.ac.in}}
\affiliation{Saha Institute of Nuclear Physics, 1/AF Bidhannagar, Kolkata-700064 (India)}
\begin{abstract}

High frequency electrostatic drift wave modes are shown as hybrid modes generated by mixing with cyclotron dynamics and grow at the expense of cyclotron modes due to mode coupling effects. The role of density gradient is to break the symmetry between the two cyclotron branches with strong values of the gradient causing the branch modified by drift to overlap with the cyclotron branch leading to a mode coupling instability. The analysis is carried out in the framework of a fluid model without invoking collisional or finite Larmor radius effects. A second order nonlinear equation with variable coefficients has been derived to govern the dynamics of high frequency electrostatic drift waves in a moving frame of (2 + 1) spatio-temporal dimensions to show the wave characteristics in various parameter ranges. The possibilities of excitation of the  high frequency electrostatic drift waves have been explored in the context of certain laboratory and astrophysical plasma systems including tokamaks and solar corona where the existence of steeper density gradients is more probable.

\end{abstract}
\maketitle
\maketitle
\section{Introduction}
Drift waves play a very prominent role in magnetically confined plasmas of various devices viz. tokamaks, Q-machines, cylindrical plasma devices etc. \cite{Weiland, Ghosh2015,Ghosh2017} including space and astrophysical plasmas viz. ionospheric plasmas, plasmas in solar atmosphere etc. \cite{Brchnelova}. The drift waves are universal modes predominantly observed in the regions of maximum density gradients in magnetized plasmas and are very crucial for transport processes \cite{Weiland}.

The research on conventional drift waves mainly focus in the low frequency frequency regime: $\omega<<\omega_{ci}$ where $\omega$ and $\omega_{ci}$ represent the drift and ion cyclotron wave frequencies respectively \cite{Horton_review}. This is because experimental devices generally have magnetic fields of the order of several kilogauss or higher. In contrast, if the magnetic field is relatively low i.e. of the order of hundreds of gauss, the possibility of electrostatic drift waves in the frequency range $\omega\geq\omega_{ci}$ cannot be avoided \cite{Ghosh2015,Ghosh2017}. In an interesting work by Klinger et al. \cite{Klingerprl} to experimentally study chaos and turbulence of drift waves, it has been shown that the drift wave fluctuation spectra may well exceed $\omega_{ci}$ where the low frequency approximation starts to fail. In our recent work on analytical modelling of the high frequency electrostatic drift waves \cite{Acharya, Acharya_Thesis} in fluid theoretical approach to derive a novel third order nonlinear equation, a detailed study of dispersion relation has been performed to explain the excitation of the high frequency electrostatic drift waves. In addition, certain solutions of the nonlinear third order differential equation have been derived with help of theory of planar dynamical systems \cite{Acharya} which results in the excitation of vortex-like structures for the high frequency electrostatic drift waves. Numerous studies have also reported the occurrence of coupled drift and ion cyclotron waves in the frequency limit: $\omega\geq\omega_{ci}$. For instance, Mikhailovskii and Timofeev \cite{Mikhailovskii} have reported a dispersion relation which represents the intersection of a drift branch and a cyclotron branch at frequencies that are multiples of $\omega_{ci}$. Hendel and Yamada \cite{Hendel} have experimentally identified ion cyclotron drift waves having frequencies
in the range: $\omega>\omega_{ci}$. The coronal heating
mechanisms caused by the drift waves in the inhomogeneous solar atmosphere has been explored by Vranjes and Poedts \cite{Vranjes2009} in the context of ion cyclotron drift waves in the limit: $\omega\approx\omega_{ci}$. A pair of nonlinear equations dealing with the coupling between electrostatic drift and ion cyclotron waves has been derived by Pokhotelov et al. \cite{Pokhotelov} in an inhomogeneous magnetoplasma. Detyna and Wooding \cite{Detyna1972} have derived a dispersion relation for ultra high frequency drift waves due to the passage of an electron stream in the region of density gradient in a plasma. They have also discussed that drifting ions in the regions of steep density gradients at the edge of a plasma column may excite unstable ultra high frequency drift waves. Instabilities of this kind may be excited in the sheath region of a low-$\beta$ plasma by an ion stream travelling at a high speed in a magnetized plasma having a background of cold stationary electrons \cite{Detyna1972}. A similar theory for high-$\beta$ plasmas also exists since similar instabilities may arise in theta-pinches or toroidal devices causing rapid plasma decay. Also, a dispersion relation has been derived in \cite{Detyna1975} for high frequency electromagnetic drift waves in the frequency limit: $\omega_{ci}<\omega<\omega_{ce}$ where $\omega_{ce}$ represents the electron cyclotron frequency due to an electron stream drifting in an inhomogeneous magnetized plasma. The linear theory for high frequency drift waves in the frequency range: $\omega\geq\omega_{ci}$ generated by the drift-cyclotron and lower-hybrid-drift instabilities has been formulated in \cite{Huba}.

Different works dealing with theoretical as well as experimental investigations showing the excitation of high frequency electrostatic drift waves in the frequency range $\omega\leq\omega_{ci} $ and $\omega>\omega_{ci}$ \cite{Ghosh2015,Ghosh2017,Klingerprl,Acharya} have inspired us to pursue a theoretical work to model these high frequency electrostatic drift waves in $2+1$ spatio-temporal dimensions. We have accomplished this with the help of a travelling wave transformation in this article. This work also happens to be an extension of our previous work \cite{Acharya} from $1+1$ to $2+1$ spatio-temporal dimensions.

This article has been organized in the following manner. The high frequency electrostatic drift waves have been studied in linear regime in Sec. \ref{lin} to derive and analyse a cubic dispersion relation. This is followed by the derivation of a second order nonlinear equation governing the dynamics of the drift waves in Sec. \ref{der}. Numerical as well as analytical solutions of the second order nonlinear equation have been explored in Sec. \ref{sol} followed by discussions of some important results of this article in Sec. \ref{res}. Finally, concluding remarks are given in Sec. \ref{con} followed by acknowledgements and bibliography.
\section{High frequency electrostatic drift waves in linear regime} \label{lin}

We consider an inhomogeneous collision-less electron-ion plasma in presence of a constant external magnetic field $B_0\hat{z}$ where $B_0$ is the magnetic field strength. The dynamics of ion fluid in this system is governed by following equations:
\begin{equation}
\frac{\partial n}{\partial t}+\vec{\nabla}.(n\vec{v})=0, \label{nlc1}
\end{equation}
\begin{equation}
\frac{\partial\vec{v}}{\partial t}+\left(\vec{v}.\vec{\nabla}\right)\vec{v}=-\vec{\nabla}\phi+\vec{v}\times\hat{z}. \label{nlc6}
\end{equation}
where Eq. (\ref{nlc1}) represent continuity equation and Eq. (\ref{nlc6}) is momentum equation. In Eqs. (\ref{nlc1}) and (\ref{nlc6}), $n,\,\vec{v}$ and $\phi$ represent ion density, ion fluid velocity and electrostatic potential respectively. As electrons obey the Boltzmann distribution, using quasi-neutrality condition, we have
\begin{equation}
n\approx n_e=n_0(x)exp\left(\phi\large \right) \label{nlc3},
\end{equation}
where $n_e$ is electron density and $n_0(x)$ denotes equilibrium ion density varying in $x$-direction.
From Eqs. (\ref{nlc1}) and (\ref{nlc3}), we get

\begin{equation}
\vec{\nabla}.\vec{v}=-\frac{d}{dt}(ln\,n_0+\phi)=-\left[\frac{\partial \phi}{\partial t}-\frac{v_x}{L_n}+\left(\vec{v}.\vec{\nabla}\right)\phi\right];\,L_n^{-1}=-\frac{d}{dx}\left(ln \,n_0\right) \label{nlc4},
\end{equation}
where $L_n$ represents density gradient scale-length. It has to be noted here that we have used following normalizations in our system:
\begin{equation}
\frac{\partial}{\partial t}\longrightarrow\frac{1}{\omega_{ci}}\frac{\partial}{\partial t};\,\vec{v}\longrightarrow\frac{\vec{v}}{c_s};\,\vec{\nabla}\longrightarrow \rho_s\vec{\nabla};\,\phi\longrightarrow\frac{e\phi}{T_e};\,n\longrightarrow\frac{n}{{n_0}};\,L_n\longrightarrow\frac{L_n}{\rho_s}, \label{nlc7}
\end{equation}
where $\omega_{ci}$ is ion cyclotron frequency, $c_s$ is ion sound speed, $\rho_s$ is ion sound radius: $\rho_s=\frac{c_s}{\omega_{ci}}$; $e$ and $T_e$ represent electronic charge and temperature respectively. The x-component of the momentum Eq. (\ref{nlc6}) is linearized by assuming $v_x,\,v_y,\,\phi$ to be proportional to $exp[i(k_xx+k_yy-\omega t)]$. Here, $k_x$ and $k_y$ refer to dimensionless wave numbers in $x$ and $y-$directions respectively whereas $\omega$ represents dimensionless frequency. These are normalized as: $k_x\rightarrow k_x\rho_s;\,k_y\rightarrow k_y\rho_s$ and $\omega\rightarrow\frac{\omega}{\omega_{ci}}$.
Therefore, we get



\begin{equation}
-i\omega v_x=-ik_x\phi+v_y \label{nlc24}.
\end{equation}
 
The linearization process is carried out in the direction of inhomogeneity by following different similar works \cite{Wakatani,Kumar,Horton_review}. The y-component of Eq. (\ref{nlc6}) gives
\begin{equation}
-i\omega v_y=-ik_y\phi-v_x \label{nlc25}
\end{equation}
Putting $v_y$ from Eq. (\ref{nlc24}) in Eq. (\ref{nlc25}), we get
\begin{equation}
v_x=\frac{1}{\omega^2-1}\left(\omega k_x+ik_y\right)\phi \label{nlc26}
\end{equation}
Similarly, we have
\begin{equation}
v_y=\frac{1}{\omega^2-1}\left(\omega k_y-ik_x\right)\phi \label{nlc27}
\end{equation}
After linearization, Eq. (\ref{nlc4}) becomes
\begin{equation}
\frac{\partial v_x}{\partial x}+\frac{\partial v_y}{\partial y}+\frac{\partial \phi}{\partial t}-\frac{v_x}{L_n}=0 \label{nlc28}
\end{equation}
This implies
\begin{equation}
ik_xv_x+ik_yv_y=i\omega\phi+\frac{v_x}{L_n} \label{nlc29}
\end{equation}
Finally, we put Eqs. (\ref{nlc26}) and (\ref{nlc27}) in Eq. (\ref{nlc29}) to get
\begin{equation}
\omega^3-(1+k^2)\omega+\frac{k_y}{L_n}-i\frac{\omega k_x}{L_n}=0 ,\label{nlc30_DR1}
\end{equation} 
which is a cubic dispersion relation to be analysed further in this section. The three exacts roots of the cubic dispersion relation Eq. (\ref{nlc30_DR1}) can be obtained with the help of Cardano's method: 

\begin{equation}
\omega_1=-p-\frac{K_1}{3p};\,\omega_2=-wp-w^2\frac{K_1}{3p};\,\omega_3=-w^2p-w\frac{K_1}{3p}, \label{r7_exroot}
\end{equation}
where $\omega_1$, $\omega_2$ and $\omega_3$ represent the three roots of Eq. (\ref{nlc30_DR1}), and, $p$ and $K_1$ are given by
\begin{equation}
p^3=\frac{K_2\pm\sqrt{K_2^2-\frac{4}{27}K_1^3}}{2};\,K_1=1+k^2+i\frac{k_x}{L_n};\,K_2=\frac{k_y}{L_n}. \label{r4}
\end{equation}

\begin{figure}
\includegraphics[height=3.1cm]{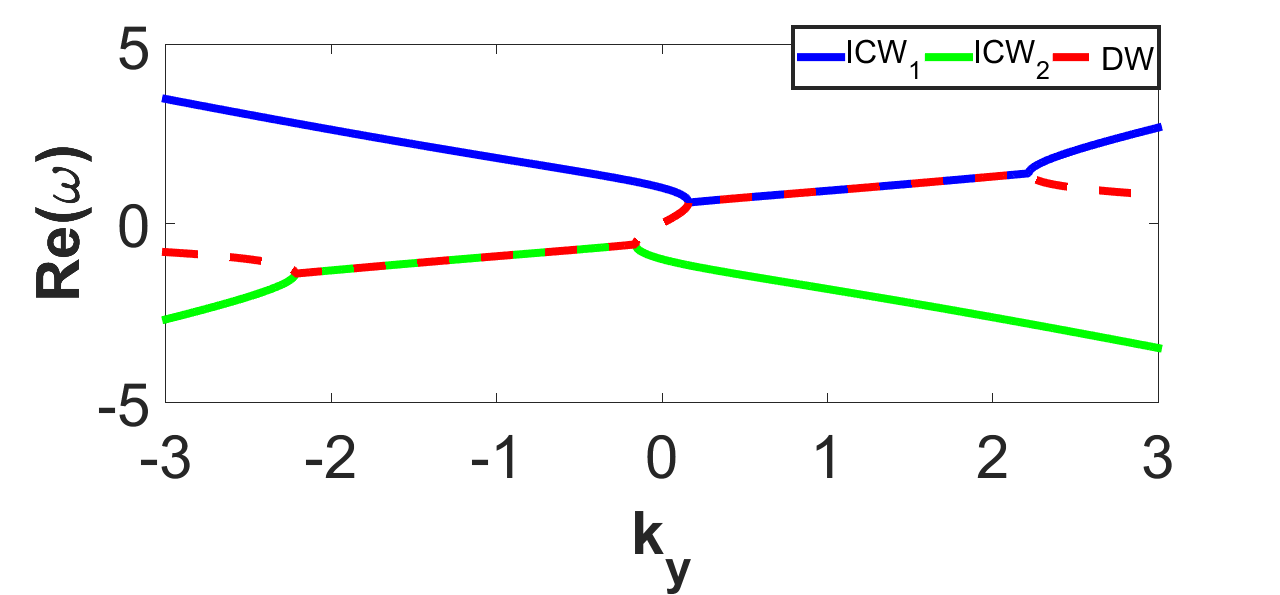} 
\
\includegraphics[height=3.1cm]{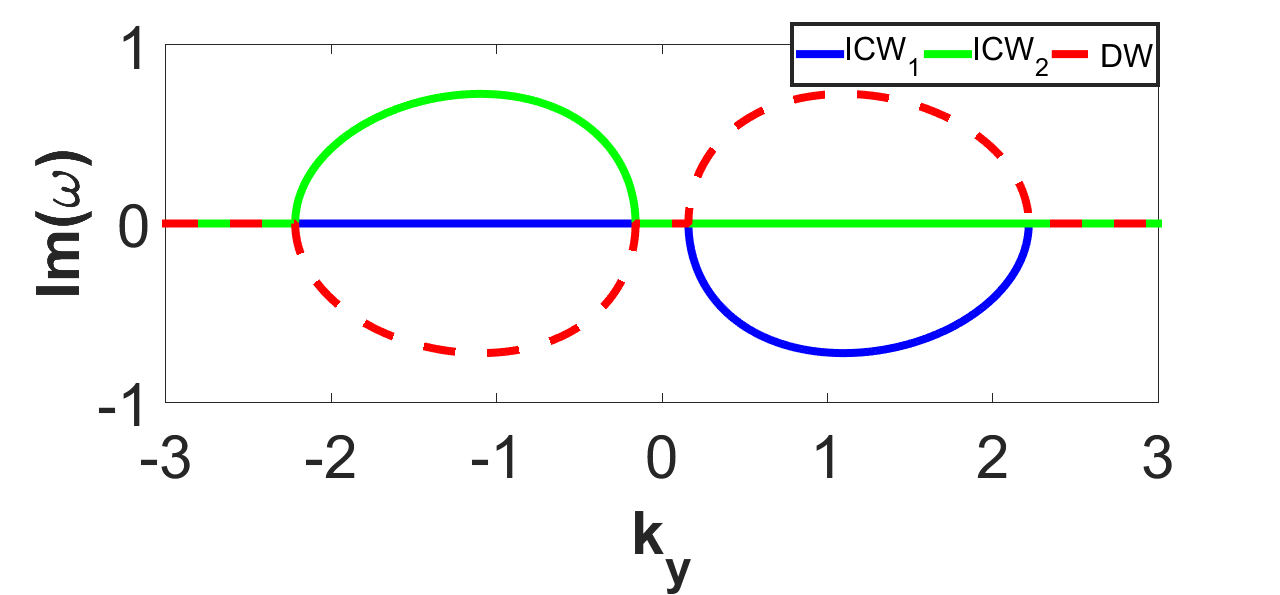}
\caption{Variations of the exact roots of the dispersion relation Eq. (\ref{nlc30_DR1}) given in Eq. (\ref{r7_exroot}) against variations in $k_y$ when $k_x=0$ and $L_n=0.4$. Overlapping regions for $Re(\omega)$ correspond to coupling of high frequency drift and ion cyclotron waves. The high frequency drift wave grows as $Im(\omega)$ is positive whereas the ion cyclotron wave damps as $Im(\omega)$ is negative.}\label{dr1}
\end{figure}

\begin{figure}
\includegraphics[height=3.1cm]{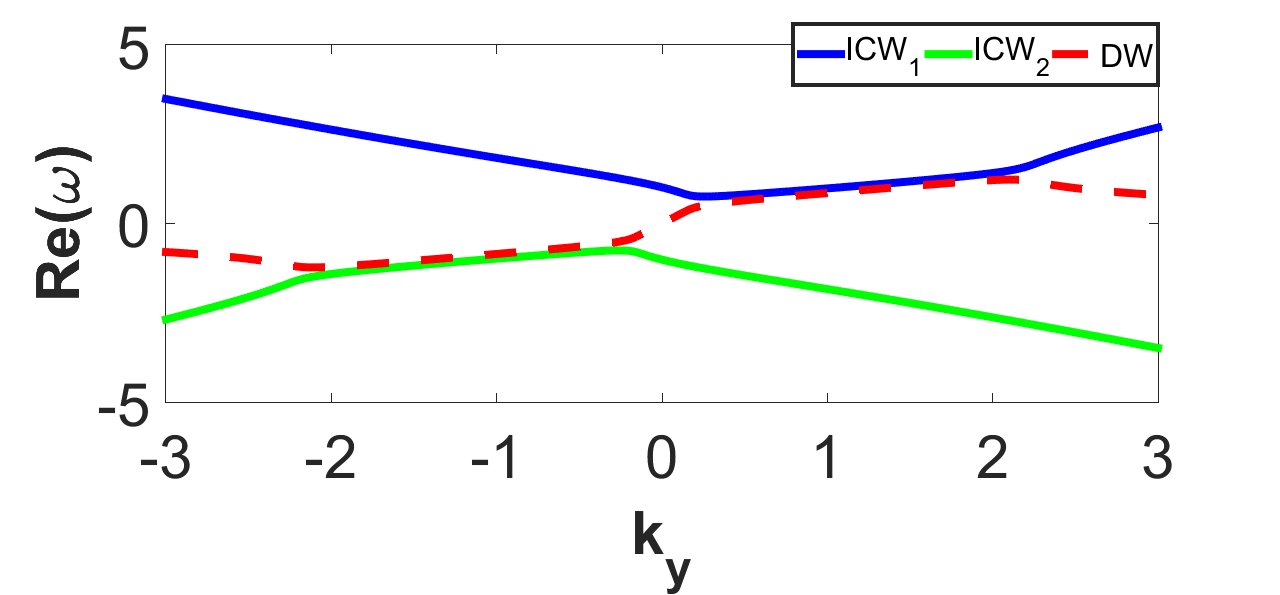}
\
\includegraphics[height=3.1cm]{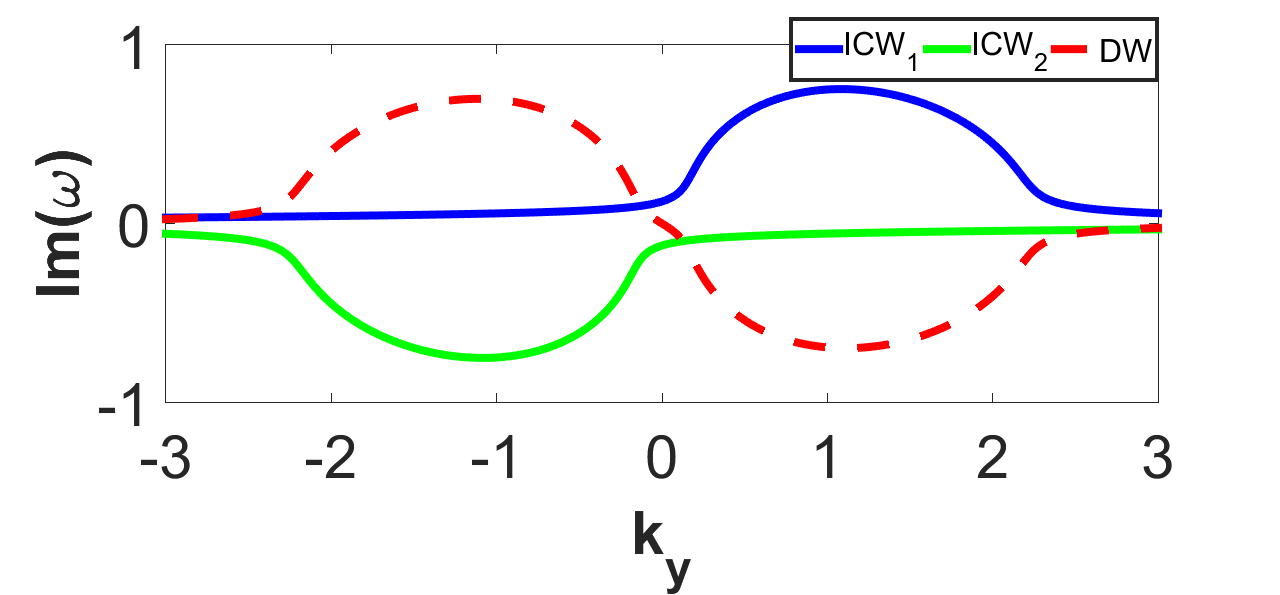}
\caption{Variations of the exact roots of the dispersion relation Eq. (\ref{nlc30_DR1}) given in Eq. (\ref{r7_exroot}) against variations in $k_y$ when $k_x=0.1$ and $L_n=0.4$. From the plot for $Re(\omega)$, it can be seen that the overlapping regions between the high frequency drift and ion cyclotron wave roots tend to separate as $k_x$ is increased.}\label{dr3}
\end{figure}



In Eq. (\ref{r7_exroot}), $w=\frac{-1+i\sqrt{3}}{2}$ is one of the three cube roots of unity having the following property:
$1+w+w^2=0;\, w^3=1$. Variations of the real and imaginary parts of the three exact roots given in Eq. (\ref{r7_exroot}) against variations in $k_y$ are plotted in Fig. \ref{dr1} for $k_x=0$ and $L_n=0.4$. The root $\omega_3$ in Eq. (\ref{r7_exroot}) corresponds to the dispersion relation for the drift wave whereas $\omega_1$ and $\omega_2$ represent ion cyclotron waves. This can be verified by carefully observing the $Re(\omega)$ plot of Fig. \ref{dr1}: when $k_y\rightarrow 0$, $Re(\omega)$ for the drift wave root corresponding to $\omega_3$ in Eq. (\ref{r7_exroot}) tends to zero whereas the remaining two ion cyclotron wave roots tend to $\omega_{ci}$. Therefore, for $k_x=0$, $\omega_3$ in Eq. (\ref{r7_exroot}) represents the drift wave whereas $\omega_1$ and $\omega_2$ represent ion cyclotron waves. The nature of the expressions for $\omega_2$ and $\omega_3$ in Eq. (\ref{r7_exroot}) implies that $\omega_2$ and $\omega_3$ are complex conjugates of each other in certain regions of the parameter space. This suggests that one of them represents the drift wave root whereas the other represents the ion cyclotron wave root. This is because the real parts of the two ion cyclotron wave roots can never be equal. In the remaining region of the parameter space, $\omega_2$ and $\omega_3$ become real and distinct. On simplifying $\omega_2$, we get
\begin{equation}
\omega_2=-\frac{1}{2}\left[\omega_1+i\sqrt{3}\left(p-\frac{K_1}{3p}\right)\right] \label{exroot1}
\end{equation}
From Fig. \ref{dr1}, it can be seen that, for $k_y>0$, $ICW_2=\omega_1$ is entirely real as $Im(\omega)$ shown in green line vanishes. In contrast, for $k_y>0$, $ICW_1=\omega_2$ is complex for certain intermediate values of $k_y$ which is shown in blue colour in Fig. \ref{dr1}. For $k_y<0$, the reverse happens: $ICW_1=\omega_2$ is entirely real whereas $ICW_2=\omega_1$ is complex for certain intermediate values of $k_y$. This fact can also be verified after some simplifications of Eq. (\ref{exroot1}). Interestingly, in contrast, $\omega_2$ in Eq. (\ref{r7_exroot}) represent the dispersion relation for the drift wave when $k_x$ is non-vanishing. When $k_x$ is increased, e.g. $k_x=0.1$ keeping $L_n$ fixed, the dispersion relation looks like Fig. \ref{dr3}. The overlapping region between the drift wave root and ion cyclotron wave roots of Eq. (\ref{nlc30_DR1}) tends to separate.


When $k_x=0$, the overlapping regions in Fig. \ref{dr1} represent the coupling or interaction between the high frequency drift and ion cyclotron waves. It can be seen in Fig. \ref{dr1} that both the drift wave and ion cyclotron wave follow a straight line in the overlapping region. This essentially means that their group and phase velocities are equal in this region. This is the requirement for the coupling to happen between the drift and ion cyclotron waves. From the plot for $Im(\omega)$ in Fig. \ref{dr1}, the drift wave is found to grow, as $Im(\omega)$ is positive, at the expense of the energy gained due to the damping of the ion cyclotron wave in the overlapping region. In contrast, when $k_x$ is non-zero, $Im(\omega)$ is negative for positive values of $k_y$ for the drift wave and vice-versa for the negative values as shown in Fig. \ref{dr3}. This essentially means that for certain intermediate values of $k_y$, the drift wave is found to damp at the expense of which the ion cyclotron wave grows. This interplay of stability behaviour between the high frequency electrostatic drift wave and ion cyclotron wave in their common overlapping region of intermediate $k_y$ values happens solely due to the density gradient and the coupling between the high frequency electrostatic drift and ion cyclotron waves. It has to be noted here that a similar work has been reported by Marchenko and Reznik \cite{Marchenko} recently where the damping of ion cyclotron wave happens due to unstable drift waves. Similar results are obtained in this manuscript where instability of the high frequency drift waves happens due to the density gradient and this drives the simultaneous damping of ion cyclotron waves. In contrary, the damping of high frequency drift waves occur at the expense of the energy spent in making the ion cyclotron waves unstable. The source of free energy driving these instabilities is solely the density gradient whose sharper values make the frequency of the drift wave high for which it gets coupled with the ion cyclotron wave only for certain intermediate values of $k_y$ affecting its growth rates. In the remaining non-overlapping regions, the high frequency electrostatic drift waves do not couple with the ion cyclotron waves. 


\subsection{$k_x=0$}
\begin{figure}
\includegraphics[width=8.4cm]{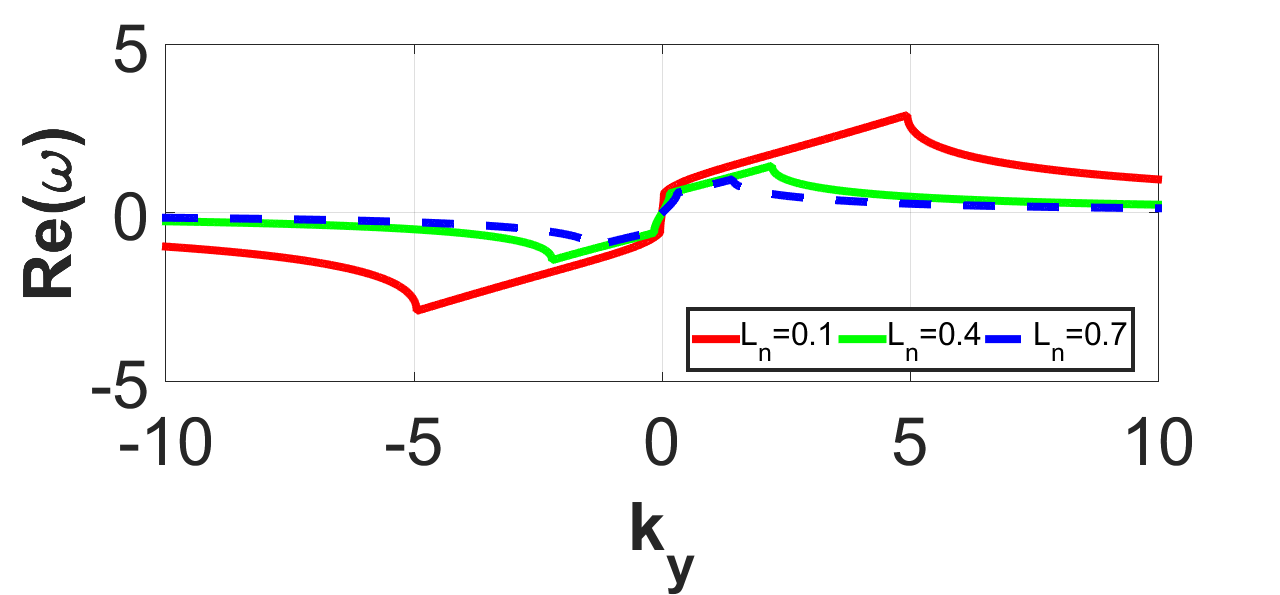}
\
\
\includegraphics[width=8.4cm]{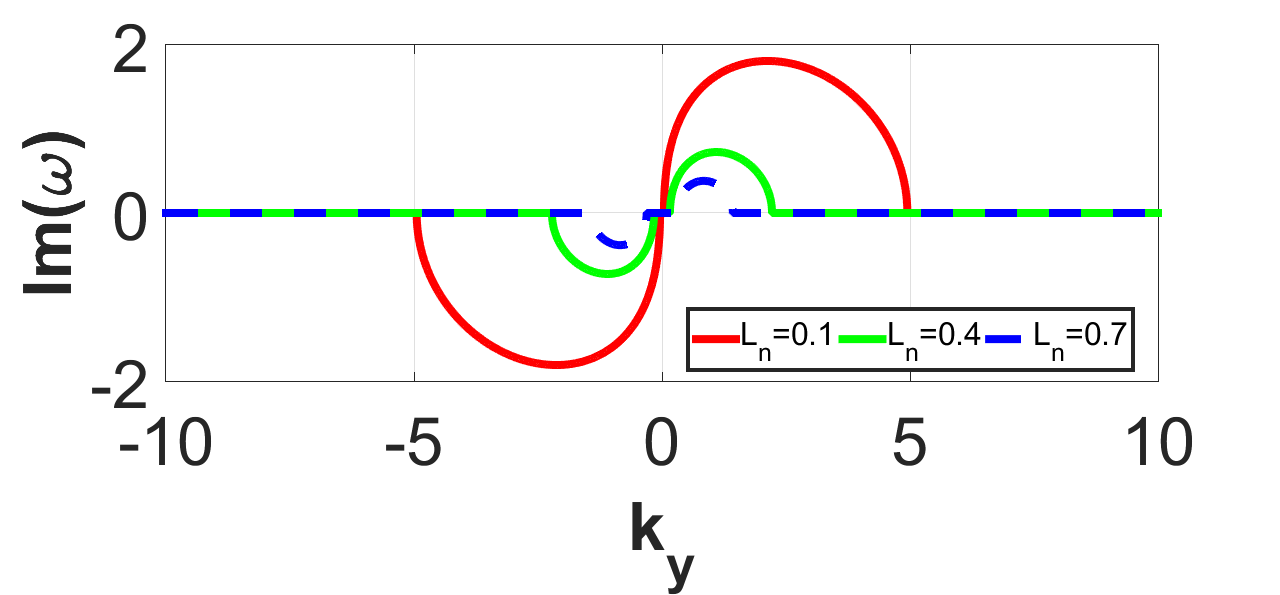}
\
\
\includegraphics[width=8.4cm]{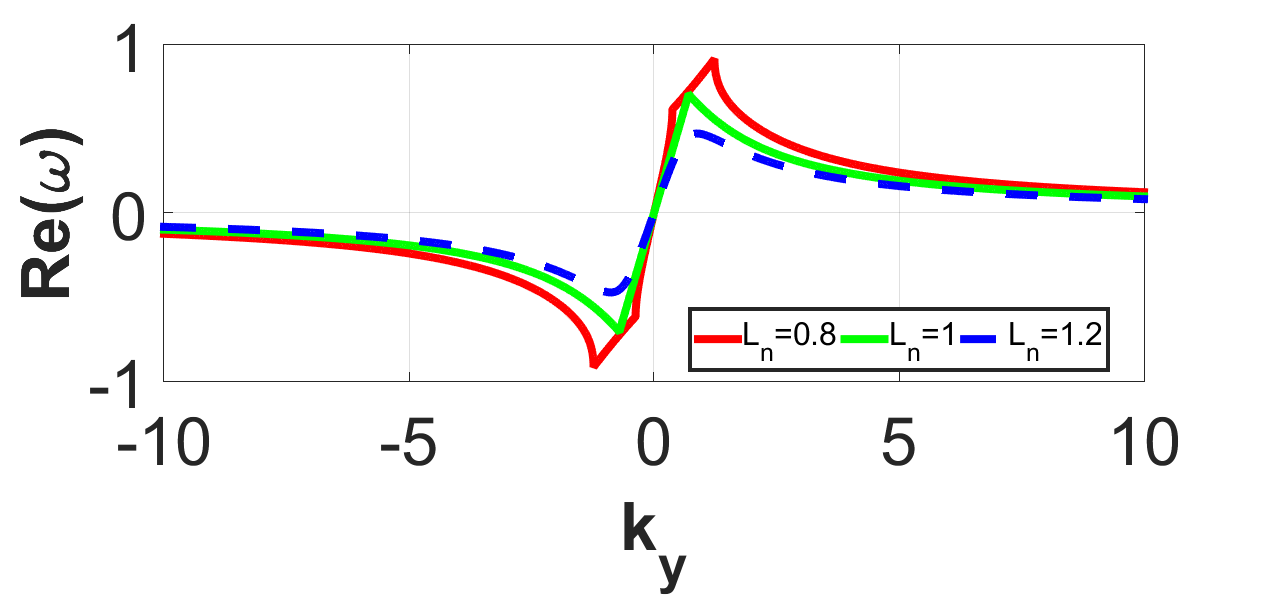}
\
\
\includegraphics[width=8.4cm]{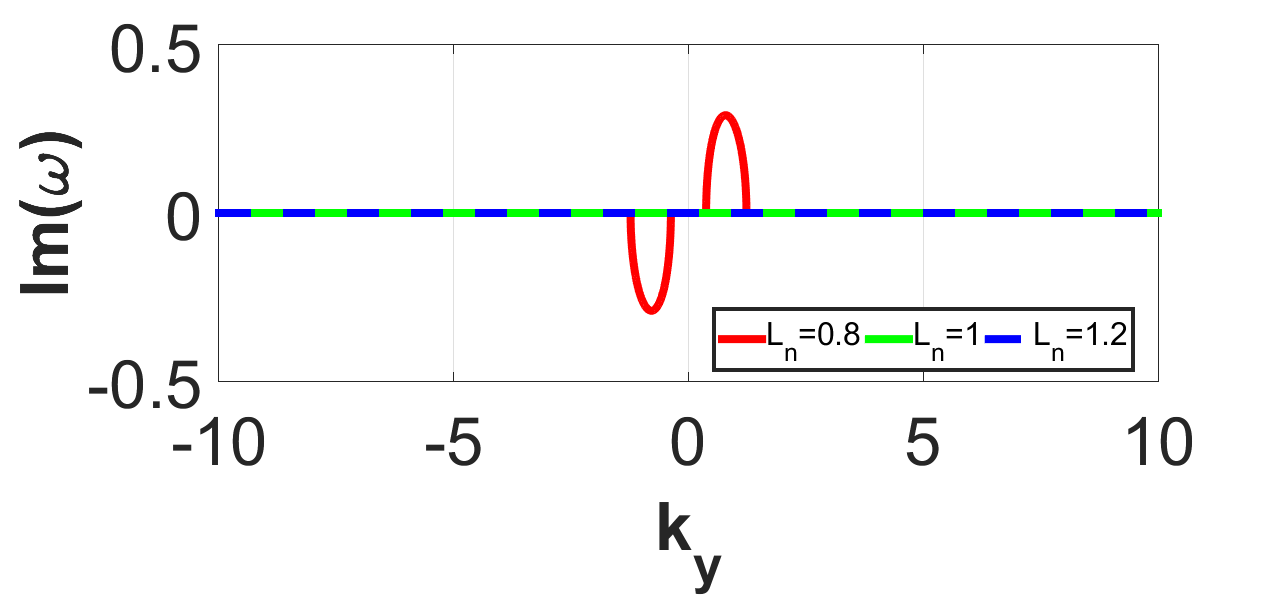}
\
\
\includegraphics[width=8.4cm]{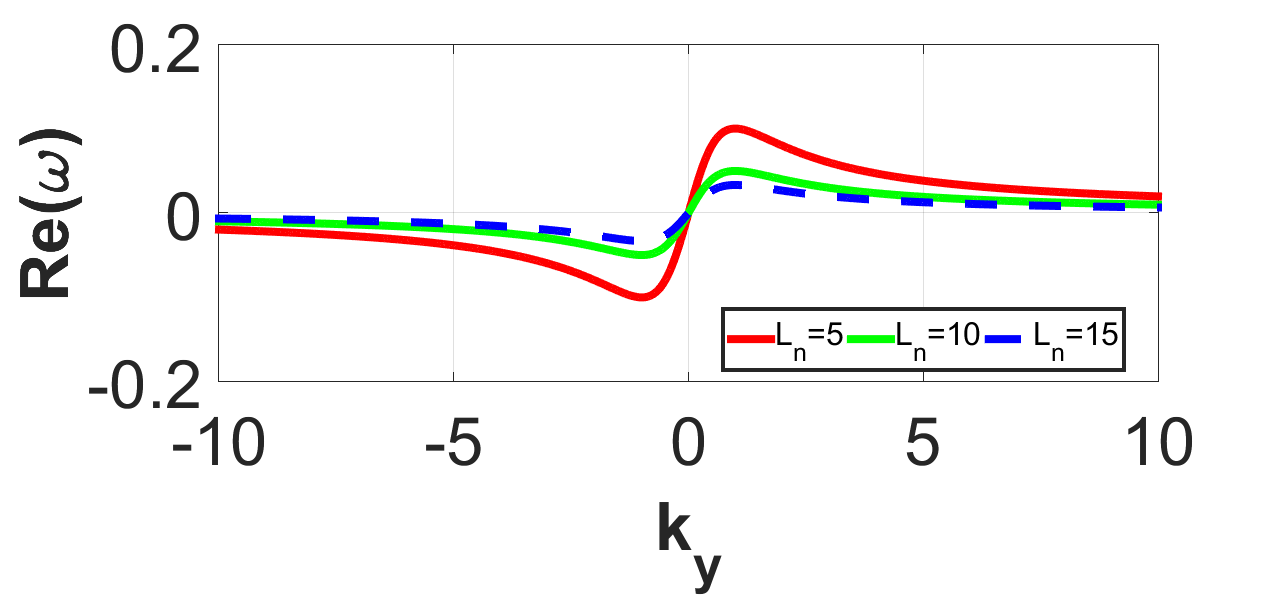}
\
\
\includegraphics[width=8.4cm]{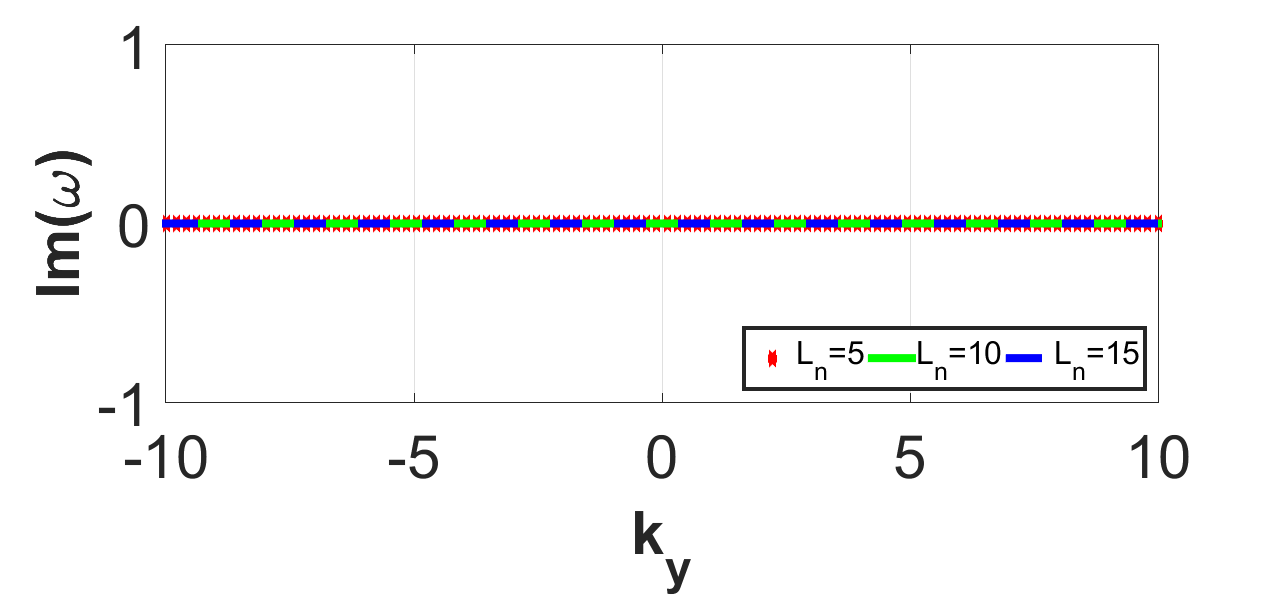}

\caption{Variations of of real and imaginary parts of the three exact roots of Eq. (\ref{nlc30_DR1}) wrt $k_y$ for different values of $L_n$ when $k_x=0$. It can be seen that extent of $k_y$-band corresponding to growth or decay rates of coupled high frequency drift and ion cyclotron waves becomes narrower as $L_n$ is increased upto $1$. For $L_n\geq 1$, $Im(\omega)$ effectively vanishes implying vanishing growth or decay rates for the drift wave when $k_x=0$. }\label{ln1}
\end{figure}

\begin{figure}
\includegraphics[height=3.5cm]{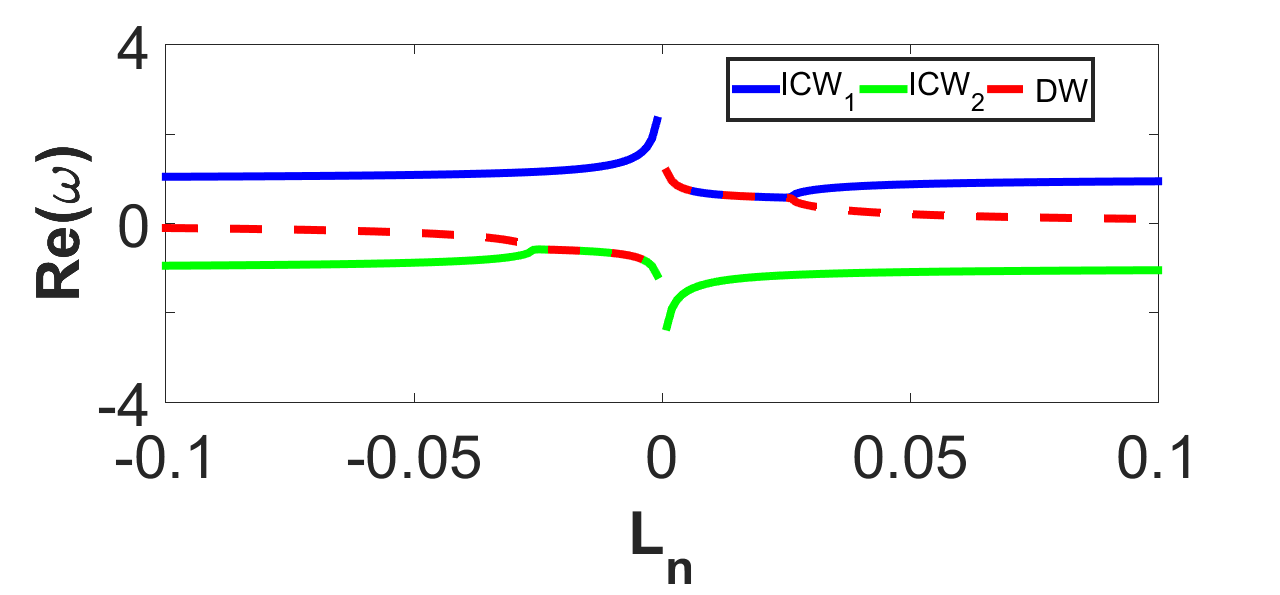}
 \ \ \ \ \includegraphics[height=3.5cm]{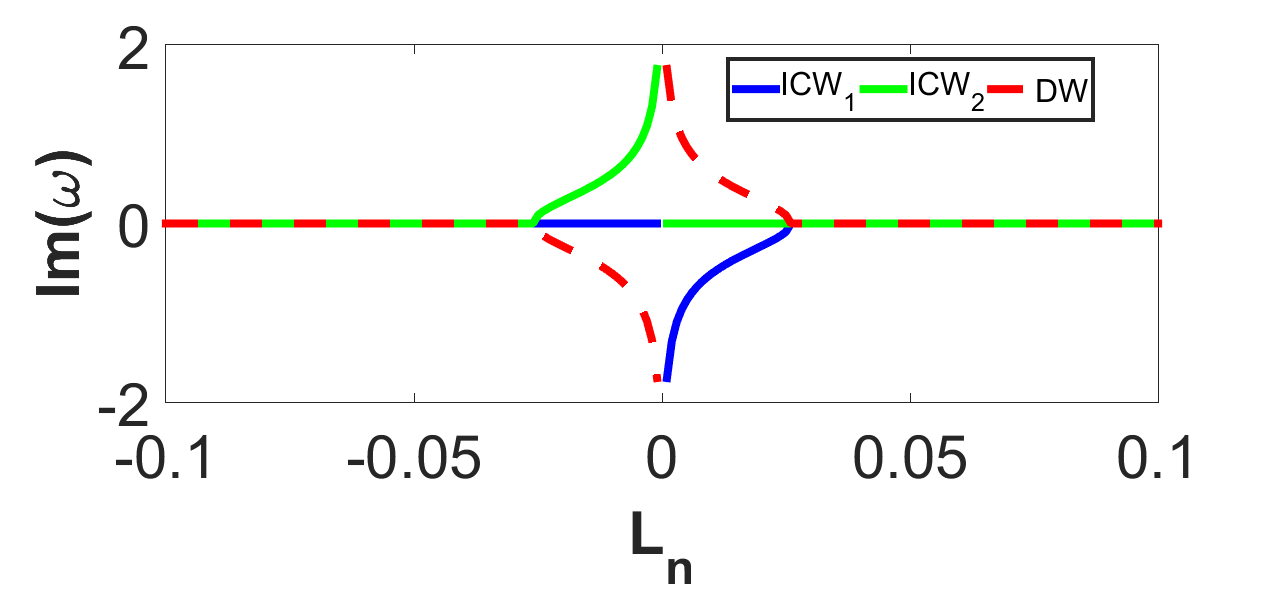}


\quad (a) $Re(\omega)$ for $k_x=0$, $k_y=0.01$ \quad\quad \qquad \qquad \quad \qquad (b) $Im(\omega)$ for $k_x=0$, $k_y=0.01$


\includegraphics[height=3.5cm]{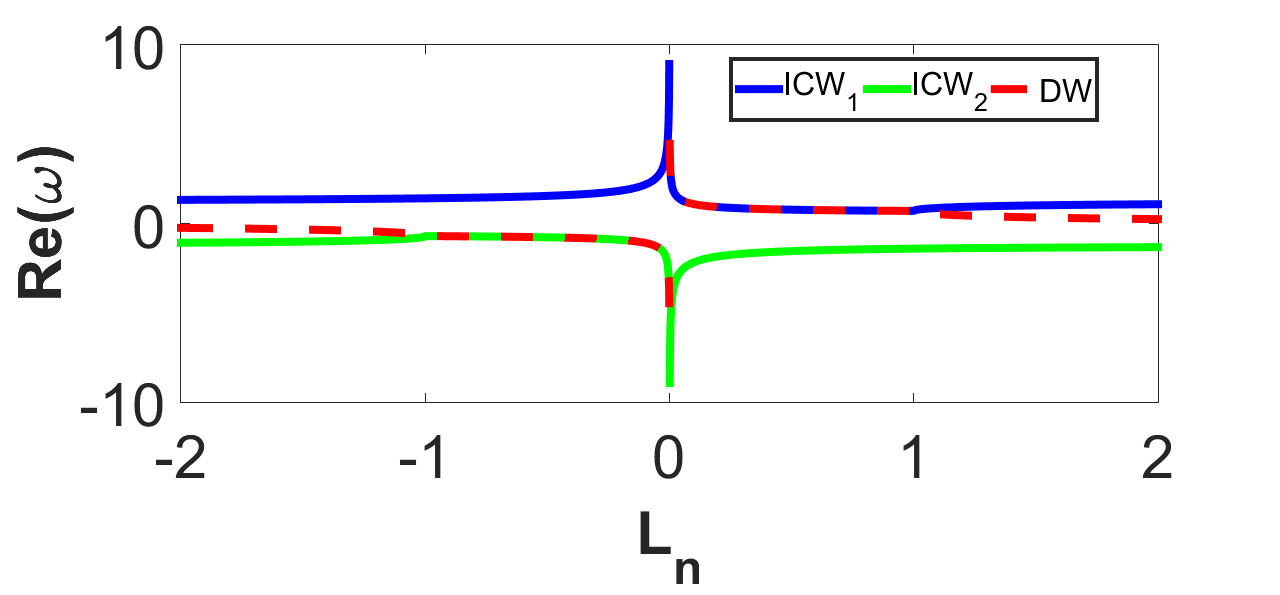}
 \ \ \ \includegraphics[height=3.5cm]{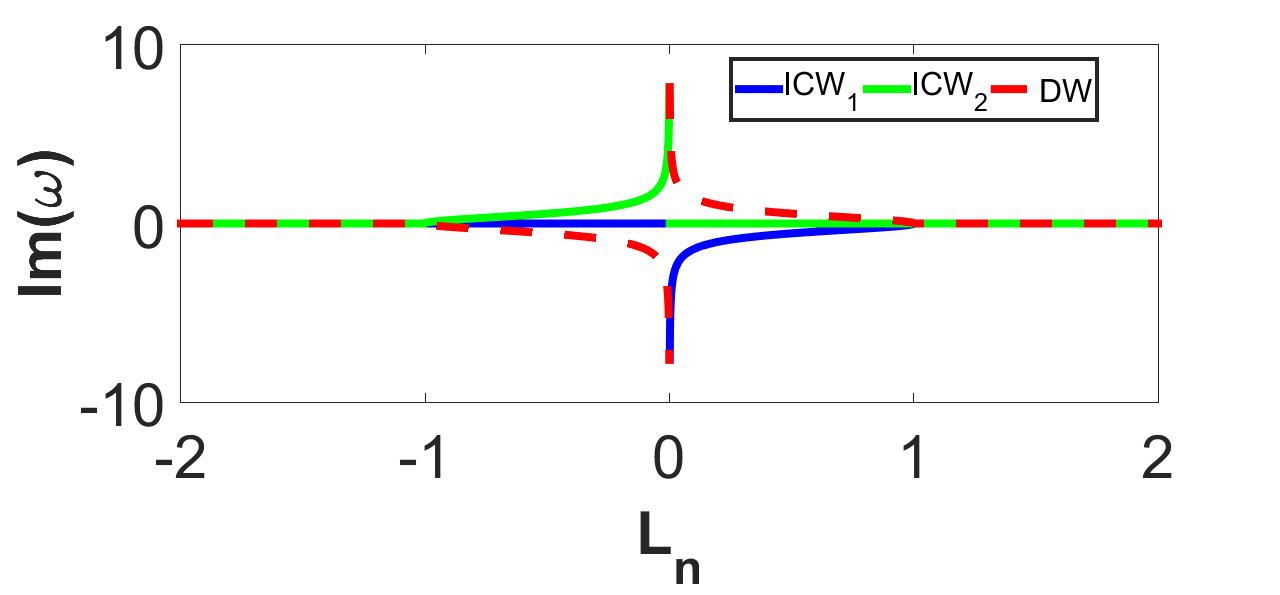}



\quad (c) $Re(\omega)$ for $k_x=0$, $k_y=0.7$ \quad\quad \qquad \qquad \quad \qquad (d) $Im(\omega)$ for $k_x=0$, $k_y=0.7$



\includegraphics[height=3.5cm]{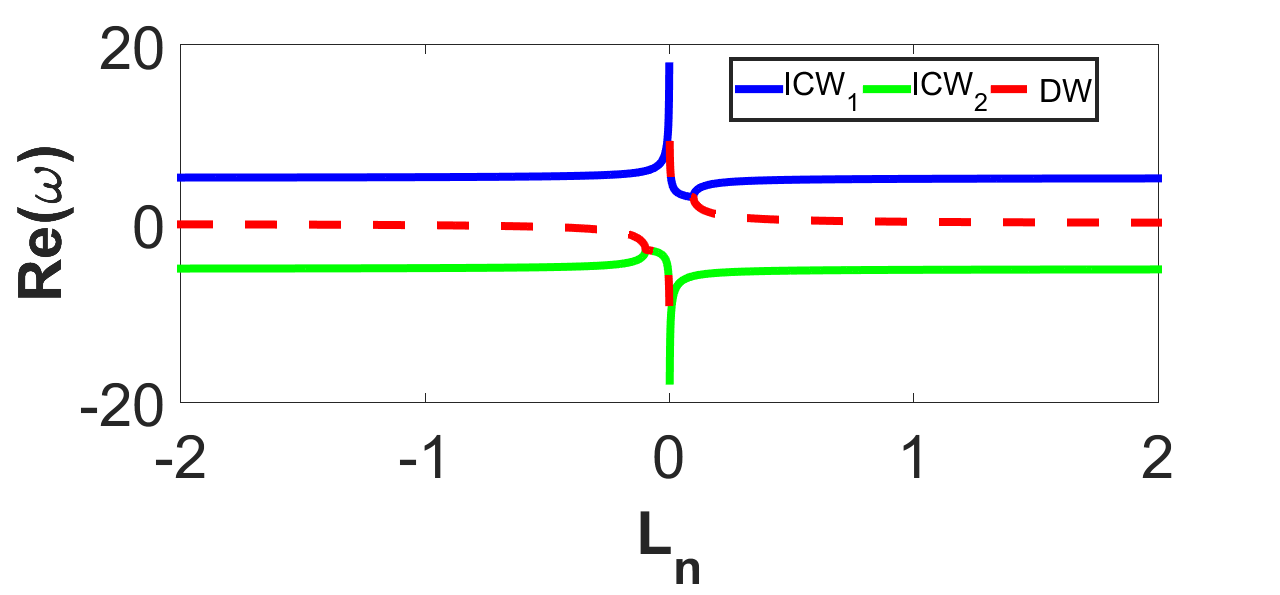}
 \ \ \ \includegraphics[height=3.5cm]{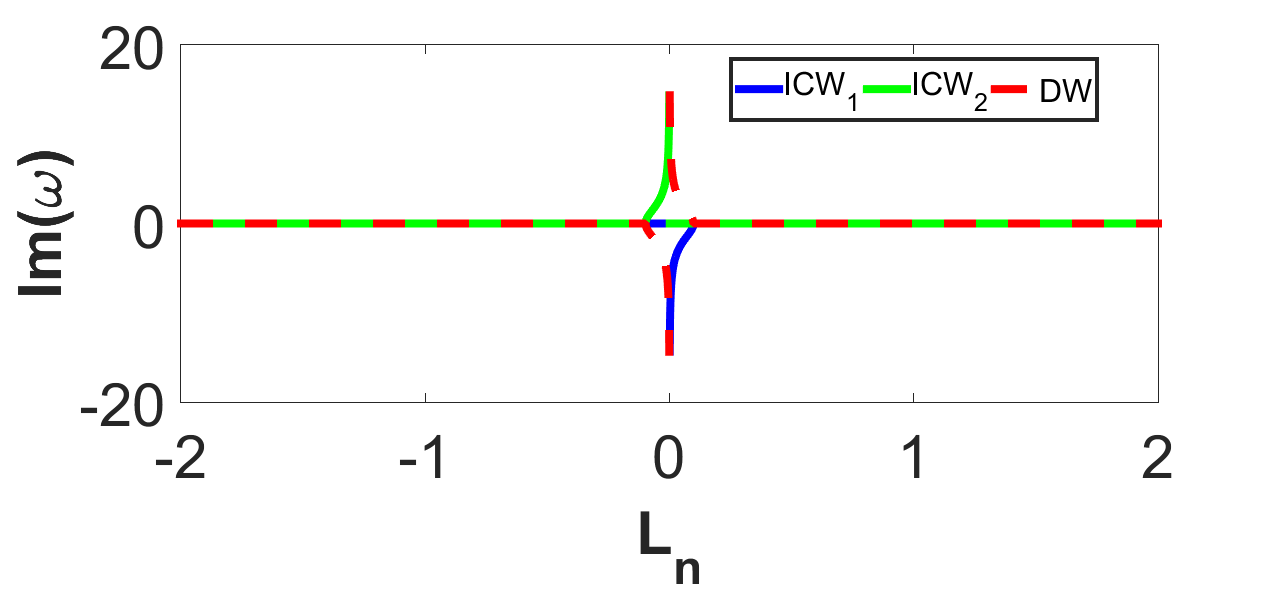}



\quad (e) $Re(\omega)$ for $k_x=0$, $k_y=5$ \quad\quad \qquad \qquad \quad \qquad (f) $Im(\omega)$ for $k_x=0$, $k_y=5$


\caption{Variations of the three exact roots of the dispersion relation Eq. (\ref{nlc30_DR1}) wrt variations in the density gradient scale length $L_n$ for different values of $k_y$. It can be noticed that the extent of the overlapping region between the high frequency drift and ion cyclotron wave roots increases upto $k_y=0.7$ after which it again decreases.}\label{lnx}
\vspace{0.5cm} 
\end{figure}

In order to explain the nature of the roots in the parameter space of $k_y$ and $L_n$ keeping $k_x=0$, we have used the discriminant $D$ of a cubic equation of the form:
\begin{equation}
ax^3+bx^2+cx+d=0, \,D=b^2c^2-4ac^3-4b^3d-27a^2d^2+18abcd. \label{disc1}
\end{equation}
There are three cases: $D>0$: three real and distinct roots; $D=0$: all real roots with at least two are equal; $D<0$: one real and two complex conjugate roots. Now the discriminant for the dispersion relation Eq. (\ref{nlc30_DR1}) is given by
\begin{equation}
D=4\left(1+k_y^2\right)^3-27\frac{k_y^2}{L_n^2} \label{disc3}
\end{equation}
 The primary effect of $L_n$ is to reduce the drift wave frequency as $L_n$ increases. It also changes the nature of the dispersion relation significantly. This can be seen from the left three sub-figures in Fig. \ref{ln1} where the real parts of the drift wave roots of Eq. (\ref{nlc30_DR1}) for different values of $L_n$ are plotted. Also, the variations of the drift wave root of the dispersion relation Eq. (\ref{nlc30_DR1}) wrt $k_y$ becomes smooth for the values $L_n>1$. For $L_n < 1$, the drift wave root shows approximately linear variations for certain intermediate values of $k_y$ and for the remaining $k_y$ values, it varies smoothly. The extent of these regions of approximately linear variations of the drift wave root tend to decrease as $L_n$ is increased. This eventually becomes a single point for $L_n=1$. Also, the overlapping of the drift and ion cyclotron wave roots disappears for $L_n=1$. From the right three sub-figures in Fig. \ref{ln1}, it can be concluded that as $L_n$ increases the range of $k_y$ values for which one ion cyclotron wave root is real whereas the other ion cyclotron wave and the drift wave roots become complex conjugates of each other becomes smaller and eventually all roots become real for $L_n\geq 1$. We have also plotted the variations of the three exact roots of the dispersion relation Eq. (\ref{nlc30_DR1}) against $L_n$ in Fig. \ref{lnx} for $k_x=0$ and $k_y=0.01;\,0.7;\,5$. From Fig. \ref{lnx}, it can be seen that $Im(\omega)\neq 0$ happens only for the overlapping region between the drift and ion cyclotron waves as before. Also, the overlapping region increases with the increase of $k_y$ upto $k_y\approx 0.7$ where the overlapping region extends upto $L_n\approx 1$. After $k_y\approx 0.7$, the overlapping region again decreases. Thus, $k_y\approx0.7$ acts as a threshold value for the growth/decay rates of drift waves. This physically signifies that when value of the density gradient scale length becomes smaller than the ion sound radius, i.e. $L_n\rightarrow \frac{L_n}{\rho_s}<1$, then, for certain intermediate values of $k_y$, one ion cyclotron wave root of the dispersion relation Eq. (\ref{nlc30_DR1}) becomes real whereas the other ion cyclotron wave root and the high frequency drift wave root become complex conjugates of each other. For the values of the density gradient scale length same as or greater than the ion sound radius, all the three roots of Eq. (\ref{nlc30_DR1}) become real.
\

\begin{figure}
\includegraphics[width=8.9cm]{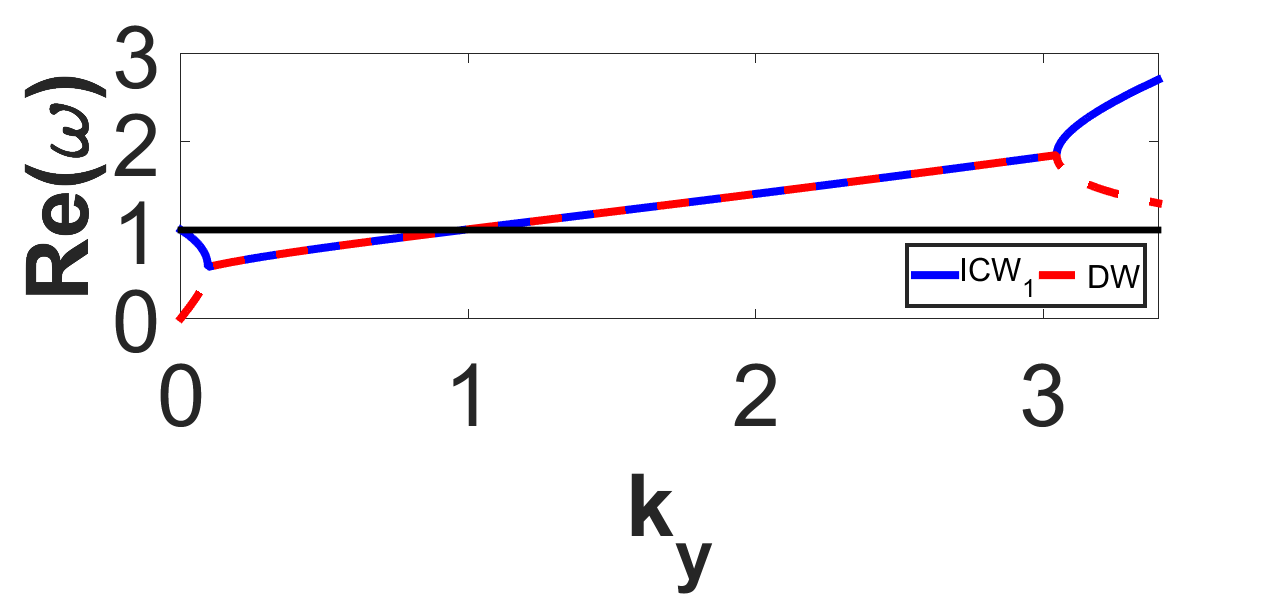}
 \includegraphics[width=8.9cm]{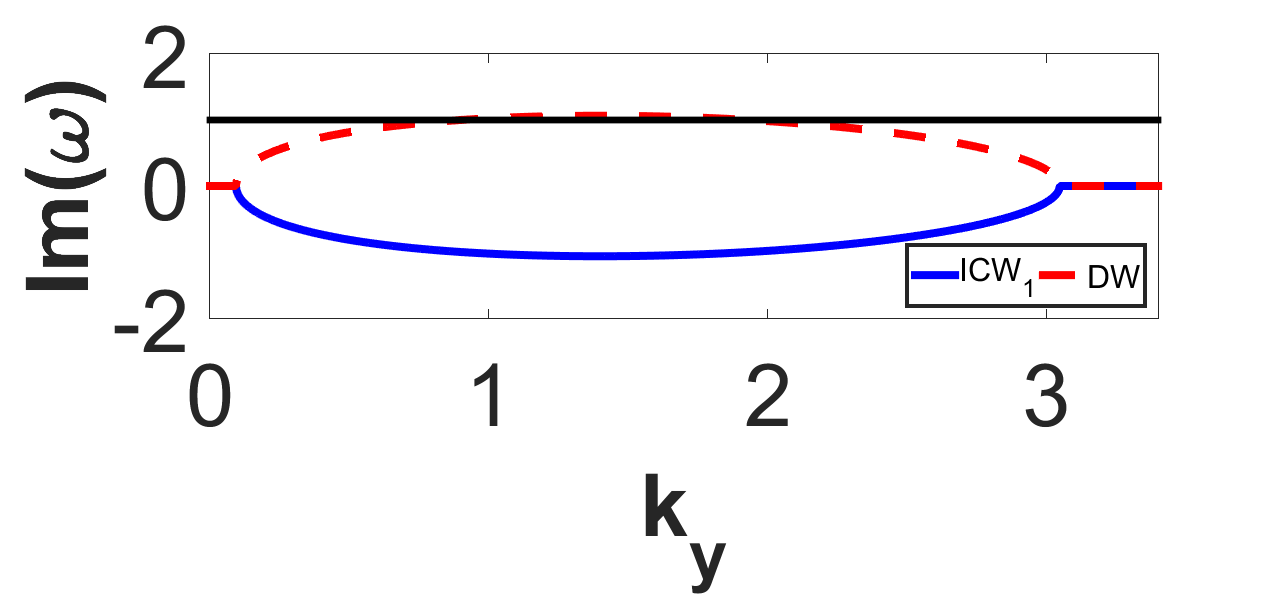}


\quad \qquad(a) Real part of $\omega$ for $k_x=0$, $L_n=0.24$ \qquad\qquad\qquad \quad \qquad (b) Imaginary part of $\omega$ for $k_x=0$, $L_n=0.24$







\caption{Behaviour of the exact roots of the dispersion relation Eq. (\ref{nlc30_DR1}) for realistic experimental values of the parameters considered in \cite{Ghosh2015}. The black line in each sub-figure corresponds to $\omega=1$ in normalized units. This is equivalent to $\omega=\omega_{ci}$. Sub-figure (a) clearly indicates that the high frequency drift wave gets coupled with the ion cyclotron wave for values of $k_y$ approximately in between $0.09$ and $3$. From sub-figure (b), it can be seen that the high frequency drift wave becomes unstable at the expense of damping of the ion cyclotron wave. This indicates a mode coupling instability wherein high frequency drift wave grows by gaining energy from damping of the cyclotron wave.}\label{real_root}
\vspace{0.5cm} 
\end{figure}

\begin{figure}
\includegraphics[width=8.4cm]{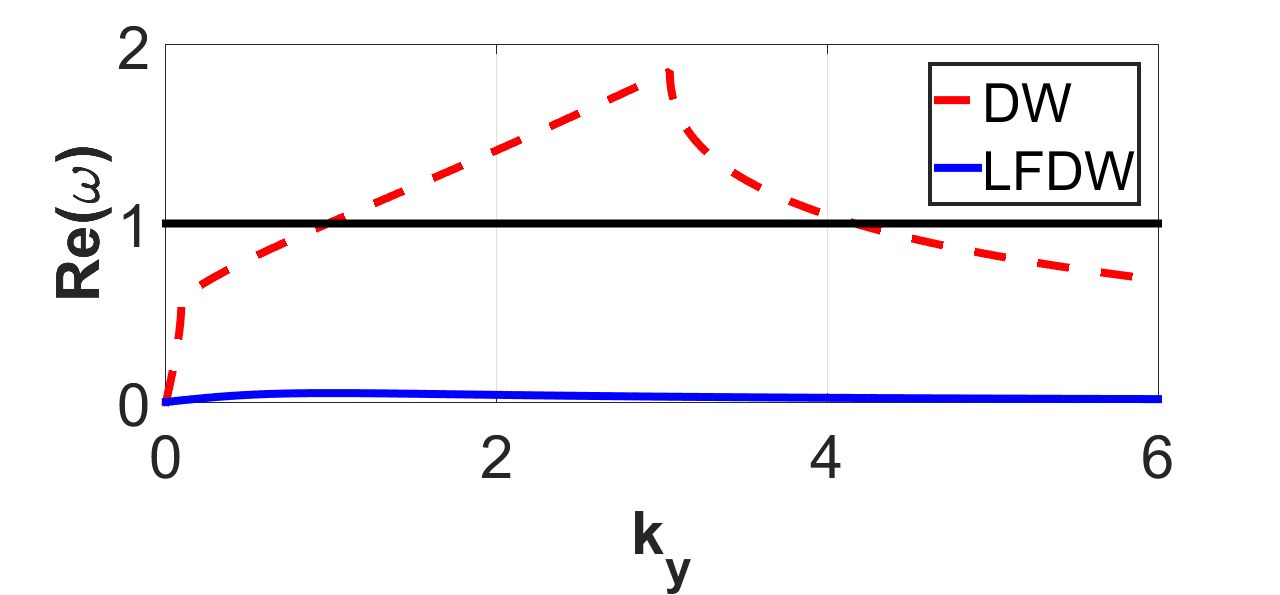}
\
\includegraphics[width=8.4cm]{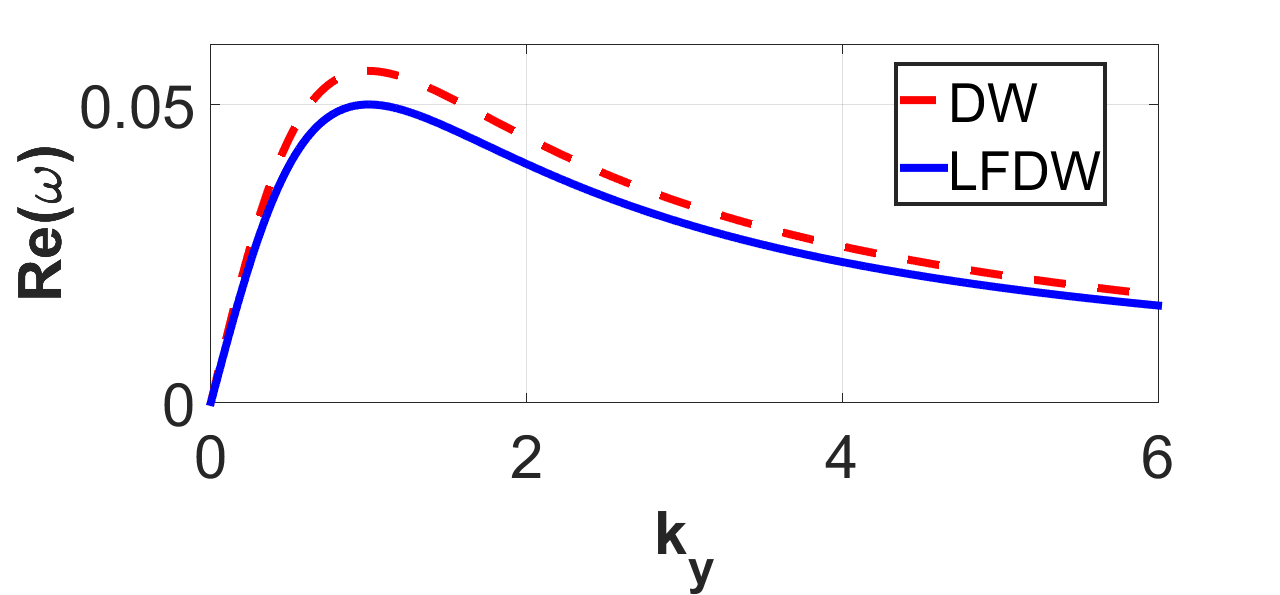}


\quad (a) $k_x=0$, $L_n=0.24$\quad\quad \qquad \qquad \qquad  \qquad \quad \qquad\qquad\qquad (b) $k_x=0$, $L_n=9$



\caption{Comparison of conventional low frequency drift wave (LFDW) for the parameters: $k_x=0$ and $L_n=10$ with the exact high frequency drift wave root of the dispersion relation Eq. (\ref{nlc30_DR1}) in sub-figure (a). Sub-figure (b) represents the same when $k_x=0$ and $L_n=9$. }\label{hmreal}
\end{figure}

We have also plotted the three roots of Eq. (\ref{nlc30_DR1}) for the realistic parameters for the excitation of high frequency drift waves as reported experimentally in \cite{Ghosh2015,Ghosh2017} in Fig. \ref{real_root}. In particular, we have considered: density gradient scale length $L_n=13\,cm$, ion sound speed $c_s=5\times10^5\,cm/s\,$ and ion gyrofrequency $\omega_{ci}=9.4\,kHz$ for sub-figures (a) and (b) in Fig. \ref{real_root}. In plotting Fig. \ref{real_root}, we have obtained the normalized density gradient scale length $L_n\rightarrow\frac{L_n}{\rho_s}=0.24$ considering the parameters taken in \cite{Ghosh2015}. From Fig. \ref{real_root}, it can be seen that $Re(\omega) \rightarrow 1$ happens in the overlapping region between the drift wave branch and ion cyclotron branch  for certain
intermediate values of $0.09<k_y<3.0$ along with $Im(\omega) > 0$ for the high frequency drift wave branch together with $Im(\omega)<0$ for the cyclotron branch. The growth of drift wave for $\omega \sim \omega_{ci} $ happens due to linear coupling between the high frequency drift mode and ion cyclotron mode. This changes the system from two independent  modes into a coupled system where one mode  becomes unstable at the expense of the other. In presence of the density gradient effects that store energy, as the high frequency drift wave branch approaches the cyclotron branch, it becomes a negative energy mode and the cyclotron branch has positive energy and their coupling leads to growth of one and damping of the other. The instability is entirely a coupling effect and happens without any finite Larmor radius effects, or collisional effects or any other kinetic
effects. The experimental investigations in \cite{Ghosh2015,Ghosh2017} where a drift mode with $\omega\geq\omega_{ci}$ is observed are supportive of this intuition. Similarly, from sub-figure (a) in Fig. \ref{hmreal}, it can be noticed that the high frequency drift wave root for $k_x=0$ and $L_n=0.24$ is very well separated from the frequency of conventional drift waves when $L_n=10$. But, when both the drift modes are compared for $L_n=9$ as shown in sub-figure (b) of Fig. \ref{hmreal}, their frequencies are not much separated. In general, the conventional low frequency drift waves have larger values of $L_n$. Only for comparison, the sub-figure (b) in Fig. \ref{hmreal} is shown.

\subsection{$k_x\neq 0$}
 As shown earlier for $k_x=0$, the high frequency electrostatic drift waves become unstable solely due to density gradient for intermediate values of $k_y$. The effects of $k_x$ on this instability is worth-investigating as $k_x$ comes with the imaginary part in the dispersion relation Eq. (\ref{nlc30_DR1}). Considering this fact, a brief study of the effects of non-vanishing values of $k_x$ on the high frequency drift waves have been explored in this subsection.
When $k_x$ is non-zero, the real part of the drift wave root tends to become distinct from the other two cyclotron wave roots as shown in Fig. \ref{dr3}. As discussed earlier, the coupling between the drift and ion cyclotron waves still happens in the straight line region due to which the drift wave root damps whereas one of the ion cyclotron wave roots grows. This is in contrast with Fig. \ref{dr1} where the high frequency drift wave grows at the expense of damping of the ion cyclotron wave root. $k_x$ is very small for drift waves and it can be neglected in many cases. Similarly, when $k_x=0$, the growth rates of the coupled drift and cyclotron waves become vanishing for $L_n\geq 1$. When $k_x$ is non-zero, the growth rates do not vanish even for $L_n\geq 1$.

In this context, the use of local analysis with a finite value of $k_x$ in the direction of the density gradient may be prone to spurious instabilities when the mode frequency gets an imaginary part proportional to $k_x$ and $L_n$ as seen in Eq. (\ref{nlc30_DR1}). Typically, these are not the true instabilities but results of the wave propagation in the direction of the density gradient. Therefore, the results with $k_x\neq 0$ should be confirmed with full nonlocal eigen-mode solutions. However, a full nonlocal analysis cannot yield meaningful results by considering the density gradient alone. Magnetic shear, temperature gradients or other inhomogeneities such as equilibrium velocity shear need to be incorporated in this regard. A full nonlocal analysis including such effects is not possible at present and will be carried out in the future.

\section{Nonliear evolution in stationary frame} \label{der}

Now we apply a stationary frame transformation of the form: 
\begin{equation}
\xi=k_xx+k_yy-\omega t;\,\psi=\omega-k_xv_x-k_yv_y;\,\alpha=k_yv_x-k_xv_y, \label{nlc11}
\end{equation}
 to the x-component and  y-component of the momentum Eq. (\ref{nlc6}) to get
\begin{equation}
\psi\frac{dv_x}{d\xi}-k_x\frac{d\phi}{d\xi}+v_y=0 \label{nlc12},
\end{equation}
and 
\begin{equation}
\psi\frac{dv_y}{d\xi}-k_y\frac{d\phi}{d\xi}-v_x=0 \label{nlc13}.
\end{equation}
 In Eq. (\ref{nlc11}), $\xi$ represents the new independent variable introduced, $\psi$ and $\alpha$ are two new dependent variables introduced which are functions of $v_x$ and $v_y$. 



Similarly, from Eq. (\ref{nlc4}), we get
\begin{equation}
\frac{d\psi}{d\xi}+\psi\frac{d\phi}{d\xi}+\frac{v_x}{L_n}=0. \label{nlc14}
\end{equation}
Now multiplying Eq. (\ref{nlc12}) with $k_x$ and Eq. (\ref{nlc13}) with $k_y$ and adding gives
\begin{equation}
\psi\frac{d\psi}{d\xi}+k^2\frac{d\phi}{d\xi}+\alpha=0;\,k^2=k_x^2+k_y^2, \label{nlc15}
\end{equation}
and multiplying Eq. (\ref{nlc12}) with $k_y$ and Eq. (\ref{nlc13}) with $k_x$ and subtracting yields
\begin{equation}
\psi\frac{d\alpha}{d\xi}-\psi+\omega=0 \label{nlc16}.
\end{equation}
Again, from Eq. (\ref{nlc11}), we have
\begin{equation}
k_yv_x-k_xv_y=\alpha, \label{nlc17}
\end{equation}
and
\begin{equation}
k_xv_x+k_yv_y=\omega-\psi, \label{nlc18}
\end{equation}
We put $v_y$ from Eq. (\ref{nlc17}) in Eq. (\ref{nlc18}) to get
\begin{equation}
v_x=\frac{1}{k^2}\left(k_y\alpha+k_x\omega-k_x\psi\right) \label{nlc19},
\end{equation}
and use Eq. (\ref{nlc19}) in Eq. (\ref{nlc14}) to obtain
\begin{equation}
\frac{d\psi}{d\xi}+\psi\frac{d\phi}{d\xi}+\frac{1}{k^2L_n}\left(k_y\alpha+k_x\omega-k_x\psi\right)=0 \label{nlc20}.
\end{equation}
Now Eqs. (\ref{nlc15}), (\ref{nlc16}) and (\ref{nlc20}) represent three first order coupled nonlinear equations in $\phi$, $\psi$ and $\alpha$. It can be seen that the linearized forms of Eqs. (\ref{nlc15}), (\ref{nlc16}) and (\ref{nlc20}) lead to the cubic dispersion relation Eq. (\ref{nlc30_DR1}). Putting $\alpha$ from Eq. (\ref{nlc15}) in Eq. (\ref{nlc16}) yields
\begin{equation}
\frac{d}{d\xi}\left(\psi\frac{d\psi}{d\xi}\right)+\frac{\psi-\omega}{\psi}+k^2\frac{d^2\phi}{d\xi^2}=0 \label{nlc21},
\end{equation}
and substituting $\alpha$ from Eq. (\ref{nlc15}) in Eq. (\ref{nlc20}) gives
\begin{equation}
\frac{d\phi}{d\xi}=\frac{1}{\frac{k_y}{L_n}-\psi}\left[\frac{d\psi}{d\xi}-\frac{k_y}{k^2L_n}\psi\frac{d\psi}{d\xi}+\frac{k_x}{k^2L_n}\left(\omega-\psi\right)\right] \label{nlc22}.
\end{equation}
Finally, substituting $\frac{d\phi}{d\xi}$ from Eq. (\ref{nlc22}) in Eq. (\ref{nlc21}), we get the final equation as follows:
$$
\left(\psi^4-\frac{k_y}{L_n}\psi^3-k^2\psi^2+\frac{k^2k_y}{L_n}\psi\right)\frac{d^2\psi}{d\xi^2}+\left(\psi^3-\frac{2k_y}{L_n}\psi^2+k^2\psi\right){\left(\frac{d\psi}{d\xi}\right)}^2+\frac{k_x}{L_n}\left(\omega-\frac{k_y}{L_n}\right)\psi\frac{d\psi}{d\xi}+\psi^3-\left(\omega+\frac{2k_y}{L_n}\right)\psi^2$$
\begin{equation}
+\frac{k_y}{L_n}\left(2\omega+\frac{k_y}{L_n}\right)\psi-\omega\frac{k_y^2}{L_n^2}=0 \label{nlc23_Finaleq}.
\end{equation}
The above Eq. (\ref{nlc23_Finaleq}) represents the final equation governing the dynamics of high frequency electrostatic drift waves in the frequency range: $\omega<\omega_{ci}$ and $\omega\geq\omega_{ci}$. This equation is identical to the nonlinear equation derived in \cite{Acharya2026} for high frequency electrostatic drift waves when the ion-neutral collision frequency $\nu$ vanishes.
\begin{figure}
\includegraphics[width=5.4cm]{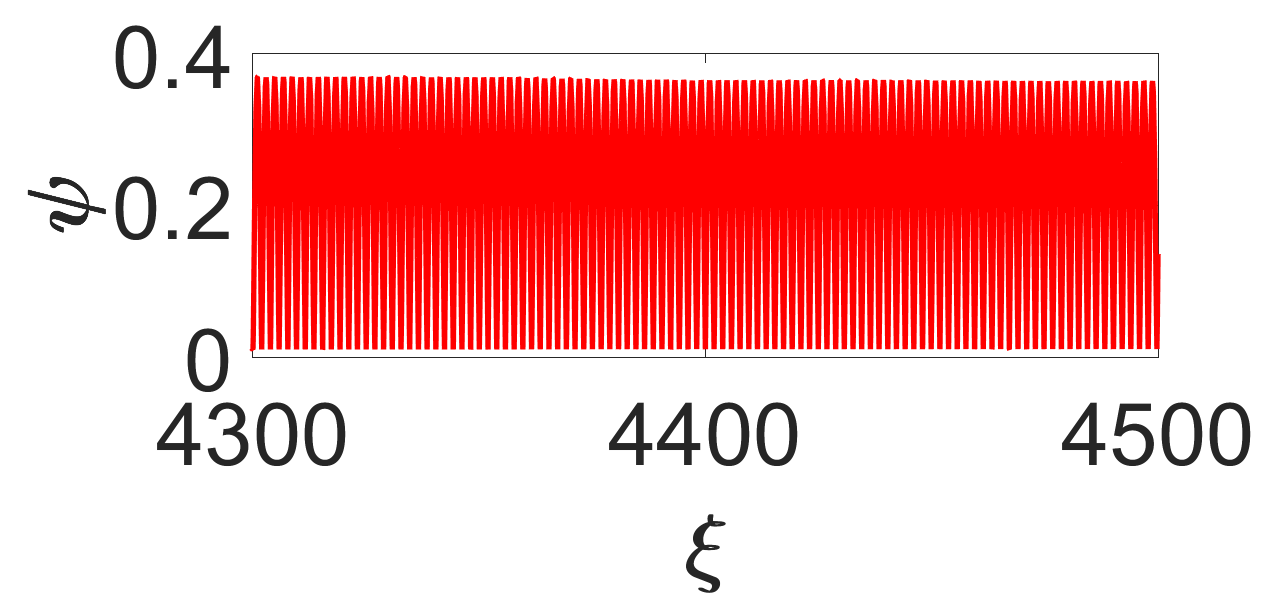}
\
\includegraphics[width=6.1cm]{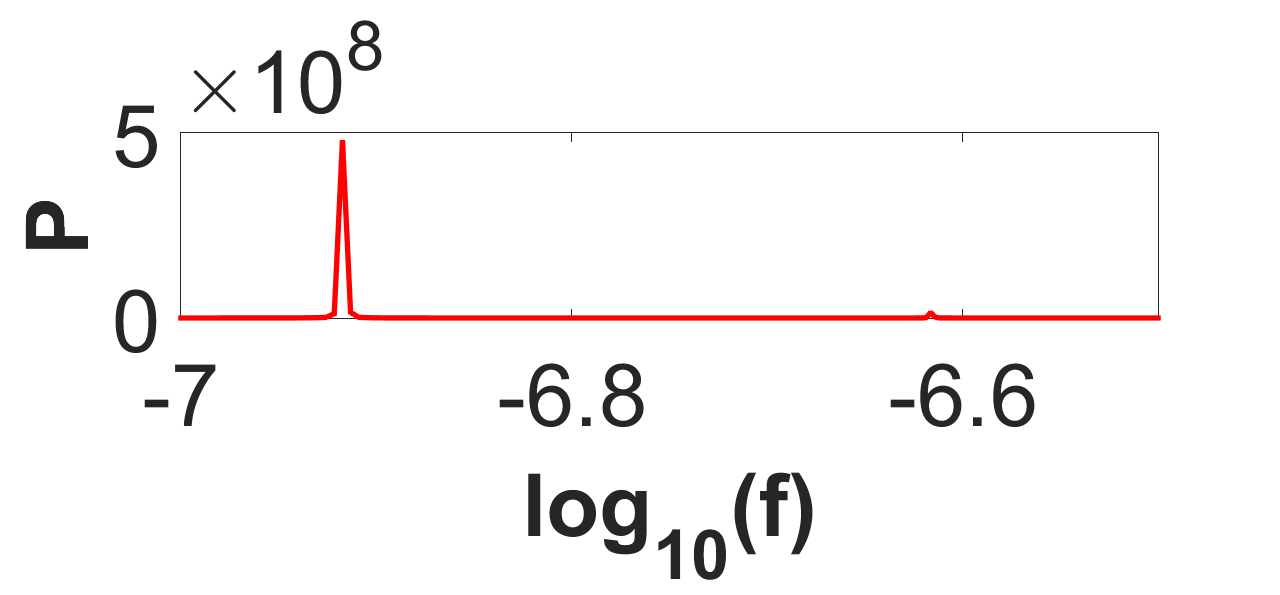}
\
\includegraphics[width=5.4cm]{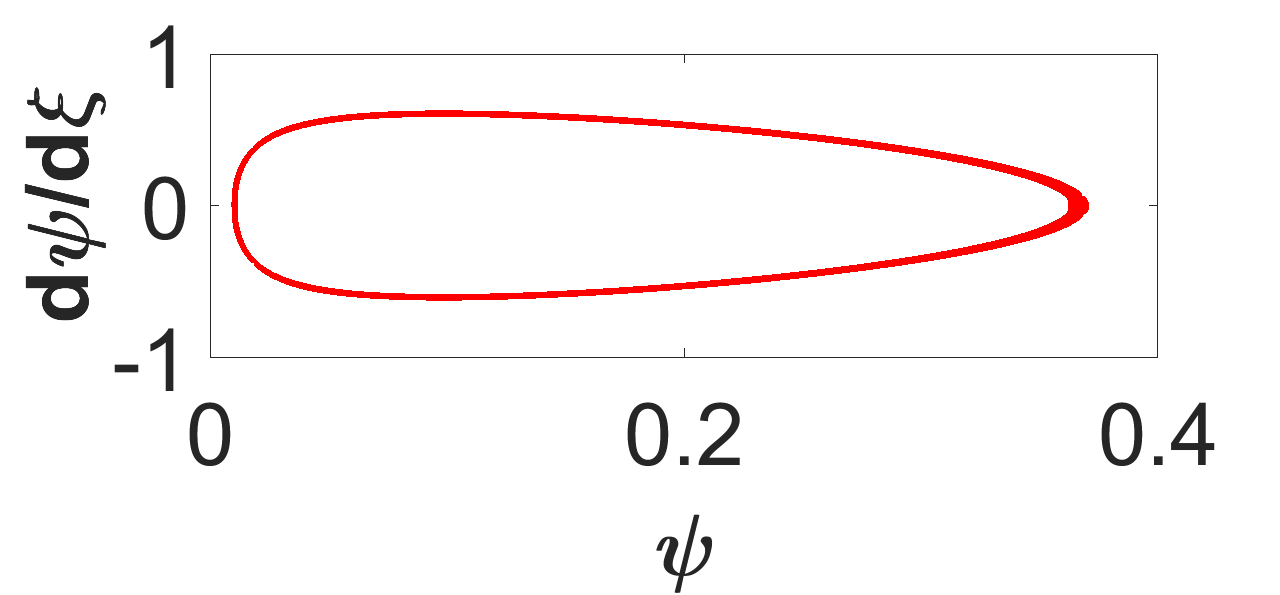}
\quad (a) Numerical solution\quad\qquad\qquad\qquad\qquad (b) Power spectrum\qquad\qquad\qquad\qquad \quad (c) Phase space 
\caption{Numerical solution of Eq. (\ref{nlc40}) for following values of the parameters: $k_x=0,\,k_y=3.1,\,\omega=0.1$ and $L_n=0.24$ which are the same as considered in Fig. \ref{real_root}. This is accompanied with its frequency domain representation and phase space. One dominant peak in the power spectral plot shown in sub-figure (b) signifies the presence of periodic oscillations in sub-figure (a). The closed trajectory of the phase space plot in sub-figure (c) also correspond to periodic oscillations.}\label{sngplott}
\end{figure}
\begin{figure}

\includegraphics[width=5.3cm]{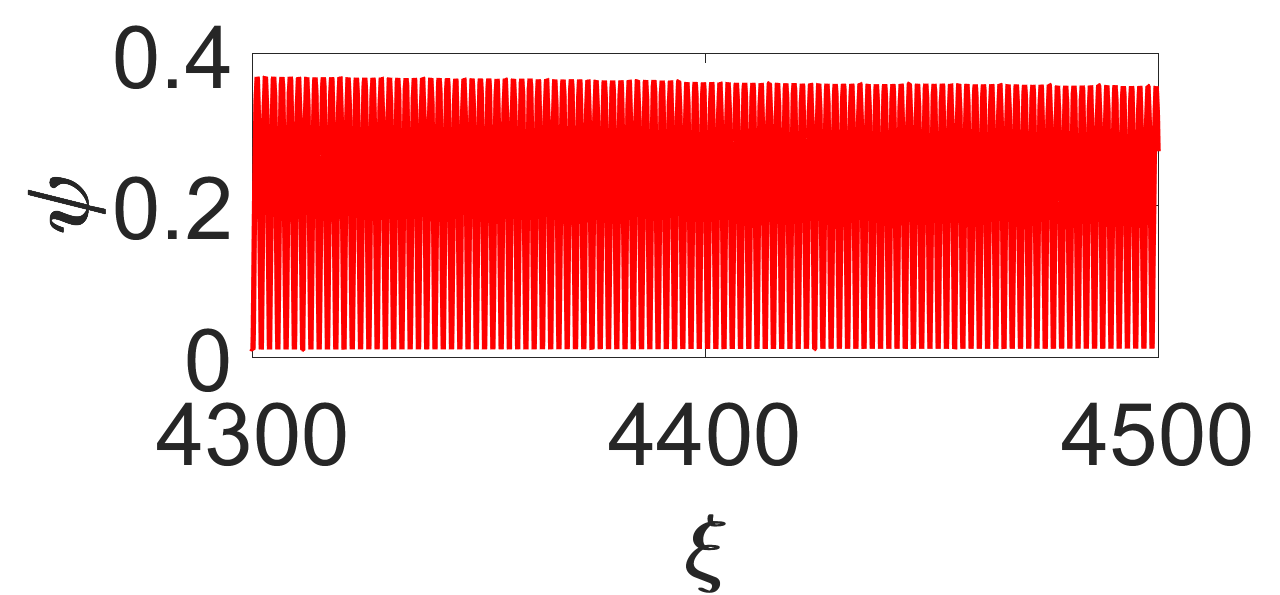}
\
\includegraphics[width=5.3cm]{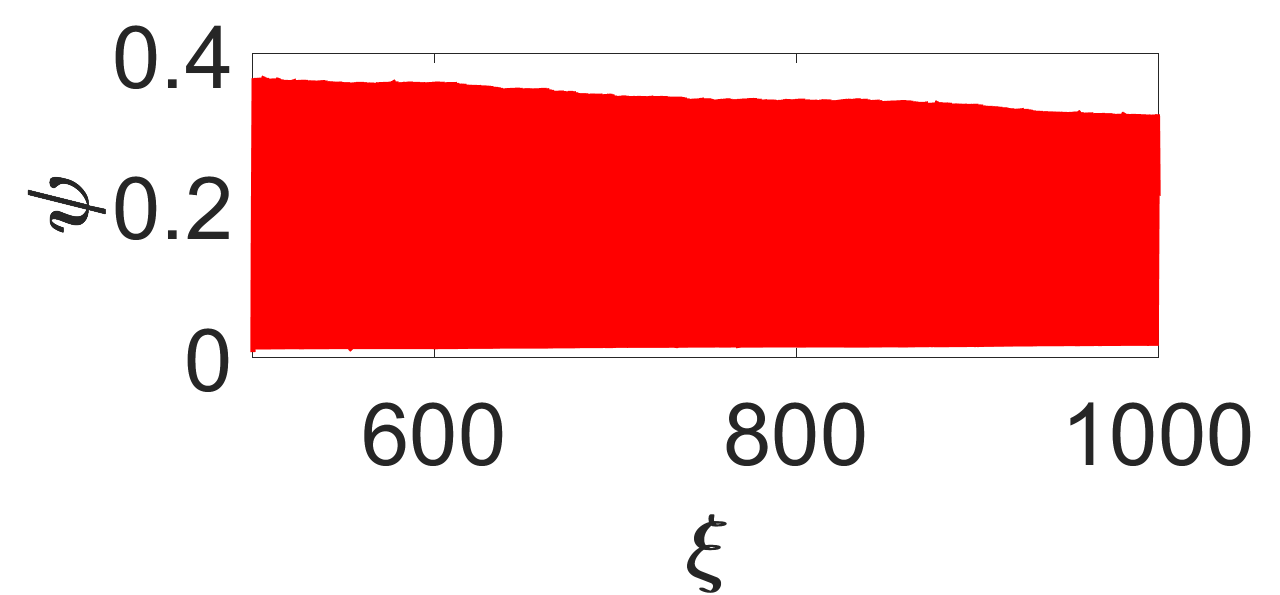}
\
\includegraphics[width=5.3cm]{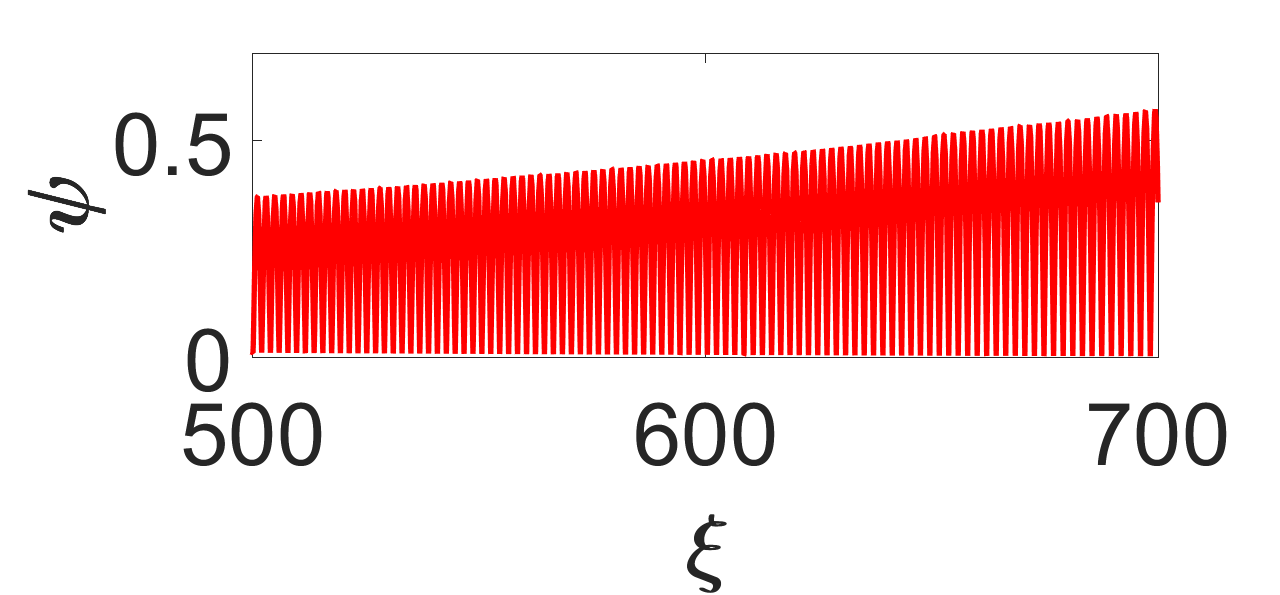}





\quad {\footnotesize (a)  $k_x=0,\,k_y=2.8,\,\omega=0.1$ and $L_n=0.24$}$\,$ $\,${\footnotesize(b) $k_x=0,\,k_y=0.9,\,\omega=0.1$, $L_n=0.4$} {\footnotesize (c) $k_x=0.01,\,k_y=3.1,\,\omega=0.1$ and $L_n=0.24$}

\caption{Numerical solutions of Eq. (\ref{nlc40}) for various values of parameters as specified in each sub-figure. periodic oscillations are clearly seen in sub-figure (a) whereas oscillations with damping and growing amplitudes can be noticed from sub-figures (b) and (c) respectively.}\label{sngplot}
\end{figure}

\section{Different solutions of second order nonlinear Eq. (\ref{nlc23_Finaleq})} \label{sol}

In this section, we have numerically as well as analytically explored different solutions of the second order nonlinear Eq. (\ref{nlc23_Finaleq}) having variable coefficients.
\subsection{Numerical Solution}
\begin{figure}
\includegraphics[width=8.4 cm]{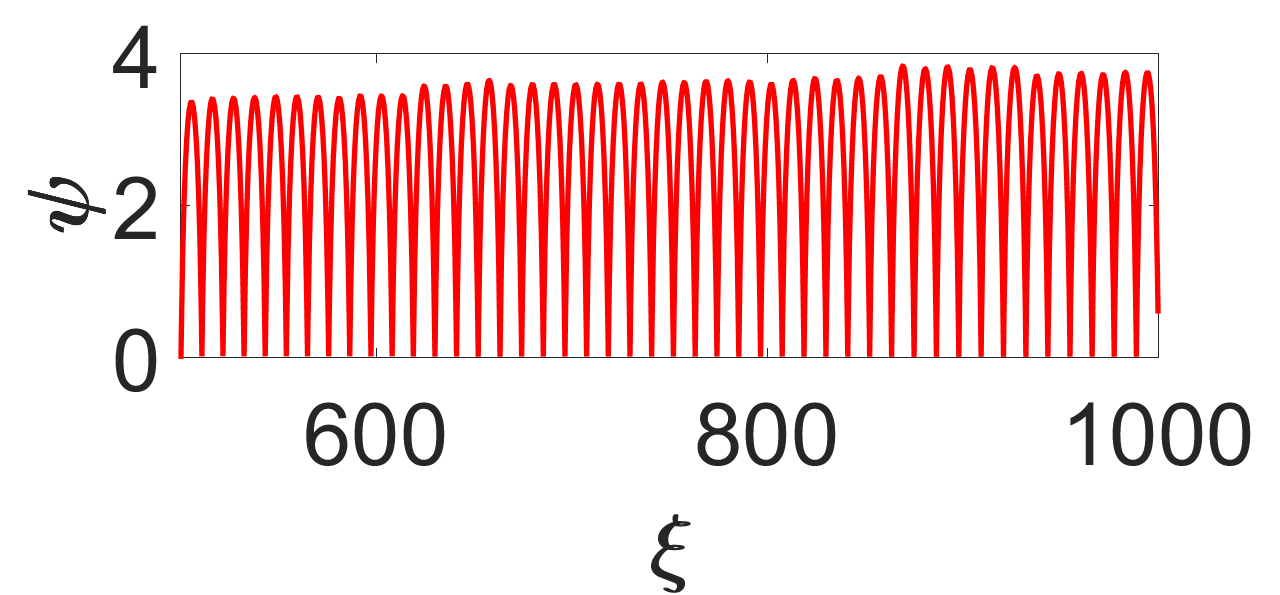}
\
\includegraphics[width=8.4 cm]{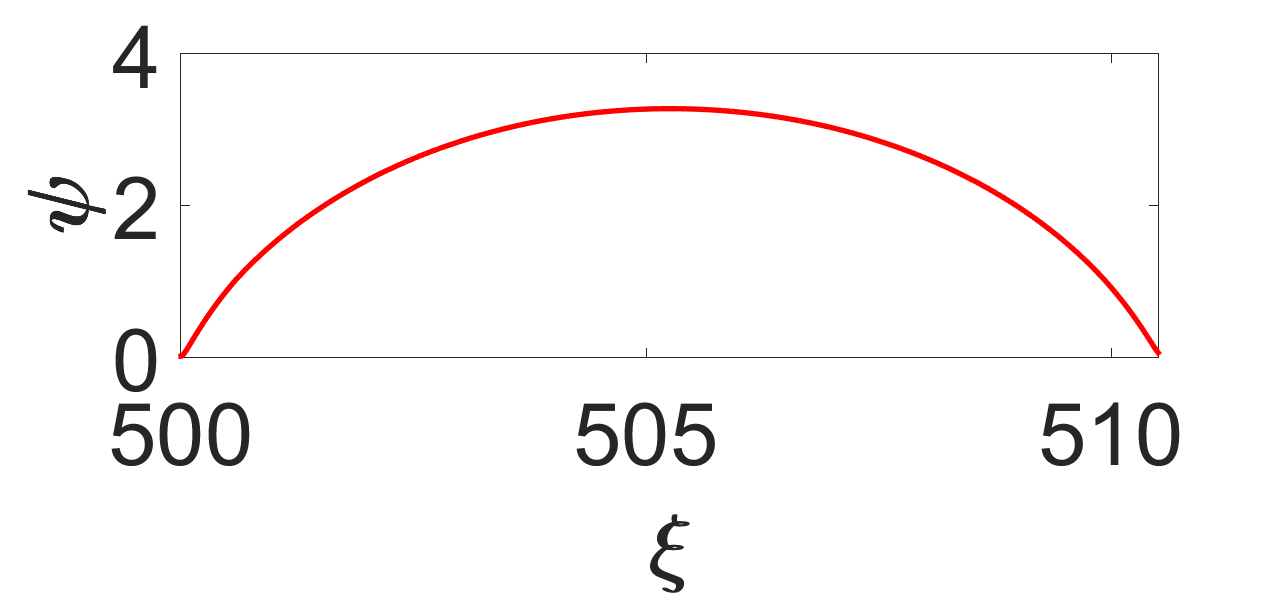}

\quad {\footnotesize (a) $k_x=0.01,\,k_y=1,\,\omega=1$, $L_n=1$} \qquad \qquad\qquad\qquad\qquad\qquad     {\footnotesize (b) $k_x=0.015,\,k_y=1,\,\omega=1$, $L_n=1$}

\caption{Nature of numerical solutions of Eq. (\ref{nlc40}) for different values of parameters. Sub-figure (a) shows typical solution of Eq. (\ref{nlc40}) with amplitude variations whereas the solution behaves like a localized pulse for the values of parameters considered in sub-figure (b).}\label{sngplot1}
\end{figure}

Now we proceed to numerically solve the second order nonlinear Eq. (\ref{nlc23_Finaleq}) in MATLAB. It can be seen that Eq. (\ref{nlc23_Finaleq}) has singularities as the coefficient of the highest order derivative in Eq. (\ref{nlc23_Finaleq}) is a function of $\psi$. Now, Eq. (\ref{nlc23_Finaleq}) has been rewritten as
$$
\frac{d^2\psi}{d\xi^2}+\frac{1}{\left(\psi-\frac{k_y}{L_n}\right)\left(\psi-k\right)\left(\psi+k\right)}
$$
\begin{equation}
\left[\left(\psi^2-\frac{2k_y}{L_n}\psi+k^2\right){\left(\frac{d\psi}{d\xi}\right)}^2+\frac{k_x}{L_n}\left(\omega-\frac{k_y}{L_n}\right)\frac{d\psi}{d\xi}+\psi^2-\left(\omega+\frac{2k_y}{L_n}\right)\psi+\frac{k_y}{L_n}\left(2\omega+\frac{k_y}{L_n}\right)-\omega\frac{k_y^2}{L_n^2}\frac{1}{\psi}\right]=0. \label{nlc40}
\end{equation}
We have solved Eq. (\ref{nlc40}) numerically and plotted the solution $\psi$ for various values of the parameters in Fig. \ref{sngplott}, Fig. \ref{sngplot} and Fig. \ref{sngplot1}. In Fig. \ref{sngplott}, the realistic values of parameters for the excitation of high frequency electrostatic drift waves taken into account in Fig. \ref{real_root} have been considered. Sub-figure (a) in Fig. \ref{sngplott} shows the typical periodic behaviour of oscillations along with its frequency representation in sub-figure (b) and phase space in sub-figure (c). From the frequency representation plot in sub-figure (b), the dimensionless dominant frequency has been found to be around $10^{-6.917}$ in addition to another peak around $10^{-6.616}$ which is approximately twice as $10^{-6.917}$. Since these two frequencies are not incommensurate, the oscillations in $\psi$ shown in Fig. \ref{sngplott} cannot be termed quasi-periodic. The periodic behaviour can also be noticed from the closed phase space plot in the sub-figure (c) of Fig. \ref{sngplott}. Now we have varied $k_x$ and $k_y$ keeping the other parameters same and shown the solutions in sub-figures (a) and (c) of Fig. \ref{sngplot}. As shown in sub-figure (a) of Fig. \ref{sngplot}, the behaviour of oscillations in $\psi$ remains similar as Fig. \ref{sngplott} when $k_y$ is varied. In contrast, if $k_x$ becomes non-vanishing as shown in sub-figure (c), amplitudes of oscillations in $\psi$ becomes continuously increased as $\xi$ becomes increased. In plotting sub-figure (b), both $k_y$ and $L_n$ are varied and the oscillations are found to show damping behaviour. It should be noted here that for the parameters considered in Fig. \ref{sngplott} and the sub-figure (a) of Fig. \ref{sngplot}, the corresponding oscillations in $\psi$ initially shows transient variations in amplitude and ultimately settles as periodic oscillations. For $k_x=0.01,\,k_y=1,\,\omega=1,\,$ and $L_n=1$ as shown in sub-figure (a) of Fig. \ref{sngplot1}, the oscillations in $\psi$ show amplitude variations. The oscillations of $\psi$ in the form of a localized pulse has been noticed when $k_x$ becomes $0.015$ as shown in sub-figure (b) of Fig. \ref{sngplot1}.

\
\subsection{Solutions based on decomposition of Eq.(\ref{nlc23_Finaleq})}
In attempt to solve Eq. (\ref{nlc23_Finaleq}) analytically, we have rewritten $\psi$ as $\psi=\omega-X$ where $X=k_xv_x+k_yv_y$. This converts Eq. (\ref{nlc23_Finaleq}) into
\begin{equation}
(A+BX)\frac{d^2X}{d\xi^2}+C{\left(\frac{dX}{d\xi}\right)}^2+D(E-X)\frac{dX}{d\xi}+FX^2+\frac{F^2}{4}X=0, \label{decs1}
\end{equation}
where $A=\omega^4-\frac{k_y}{L_n}\omega^3-k^2\omega^2+\frac{k^2k_y}{L_n}\omega$; $B=-4\omega^3+\frac{3k_y}{L_n}\omega^2+2k^2\omega-\frac{k^2k_y}{L_n}$; $C=-\omega\left(\omega^2-\frac{2k_y}{L_n}\omega+k^2\right)$; $D=\frac{k_x}{L_n}\left(\omega-\frac{k_y}{L_n}\right)$; $E=\omega$; $F=-2\left(\omega-\frac{k_y}{L_n}\right)$. In Eq. (\ref{decs1}), third and higher order terms have been neglected. It has to be noted here that the parameter $D$ considered in this subsection is different from that considered earlier in Sec. \ref{lin}.

\begin{figure}
\includegraphics[width=15.4cm]{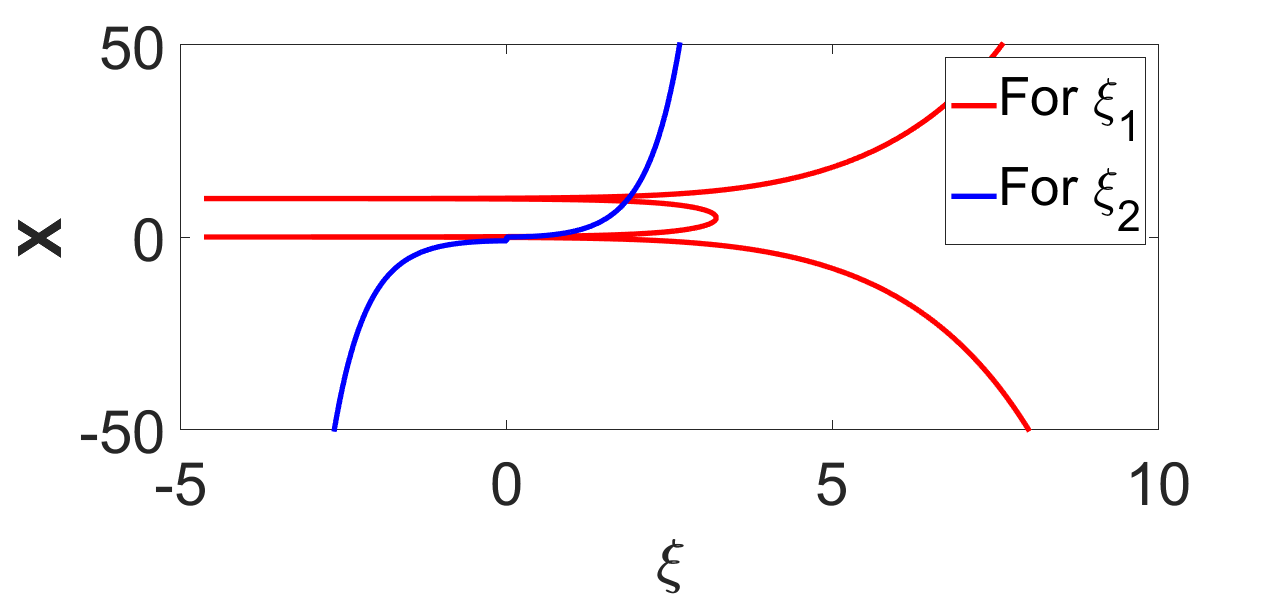}
\caption{Pictorial representation of the nature of exact solutions Eqs. ({\ref{decs5a}}) and (\ref{decs6a}) for the following values of the parameters: $A'=B'=D'=E'=1$ and $C'=10$.}\label{ad1}
\end{figure}

Now we decompose Eq. (\ref{decs1}) into two separate parts as given below:
\begin{equation}
\left[\frac{d}{d\xi}\left(A\frac{dX}{d\xi}+BX\frac{dX}{d\xi}+DEX-\frac{D}{2}X^2\right)\right]+\left[(C-B){\left(\frac{dX}{d\xi}\right)}^2+FX^2+\frac{F^2}{4}X\right]=0. \label{decs2}
\end{equation}
Here, for simplicity to get solutions of the second order Eq. (\ref{decs1}) by reducing it into two first order equations, we have assumed that the terms in Eq. (\ref{decs2}) separated by square brackets vanish separately. This results in

\begin{equation}
A\frac{dX}{d\xi}+BX\frac{dX}{d\xi}+DEX-\frac{D}{2}X^2=0, \label{decs3} 
\end{equation}
and
\begin{equation}
(C-B){\left(\frac{dX}{d\xi}\right)}^2+FX^2+\frac{F^2}{4}X=0 \label{decs4}.
\end{equation}
Such a decomposition is similar to that done in \cite{Acharya2026}. Now both Eqs. (\ref{decs3}) and (\ref{decs4}) can be solved analytically. Eq. (\ref{decs3}) gives the solution:
\begin{equation}
\xi_1=-\frac{A}{DE} ln \,(X)+\left(\frac{A}{DE}+\frac{2B}{D}\right)ln\,(X-2E)+K_1, \label{decs5}
\end{equation}
and Eq. (\ref{decs4}) yields:
\begin{equation}
\xi_2=2\sqrt{\frac{B-C}{F}}ln\left\lbrace sec\left[tan^{-1}\left(\sqrt{\frac{4X}{F}}\right)\right]+\sqrt{\frac{4X}{F}}\right\rbrace+K_2, \label{decs6}
\end{equation}
where $K_1$ and $K_2$ represent the constants of integration. As the expressions in Eqs. (\ref{decs5}) and (\ref{decs6}) represent the exact solutions of Eqs. (\ref{decs3}) and (\ref{decs4}) respectively, the intersections of these two exact solutions represent the desired solutions of the second order Eq. (\ref{decs1}) for the constraint that the two expressions in the square brackets of Eq. (\ref{decs2}) vanish separately. We have plotted the typical nature of the solutions in Fig. \ref{ad1} for arbitrarily selected values of parameters. For plotting Fig. \ref{ad1}, we have rewritten the solutions Eqs. (\ref{decs5}) and (\ref{decs6}) in the following form:
\begin{equation}
\xi_1=A' ln(X)+B' ln(X-C')+K_1 \label{decs5a},
\end{equation}
and
\begin{equation}
\xi_2=D'ln\left\lbrace sec\left[tan^{-1}\left(\sqrt{E'X}\right)\right]+\sqrt{E'X}\right \rbrace+K_2, \label{decs6a}
\end{equation}
and taken $A'=B'=D'=E'=1$ and $C'=10$. The integration constants $K_1$ and $K_2$ are assumed to be vanishing. From Fig. \ref{ad1}, it can be seen that the high frequency drift wave propagate in the form of localized pulses which are similar to modulated solitary waves \cite{Yu}.

\subsection{Derivation of an exact solution in asymptotically small limit of $\psi$: $\psi<<1$}
\begin{figure}
\includegraphics[width=8.4cm]{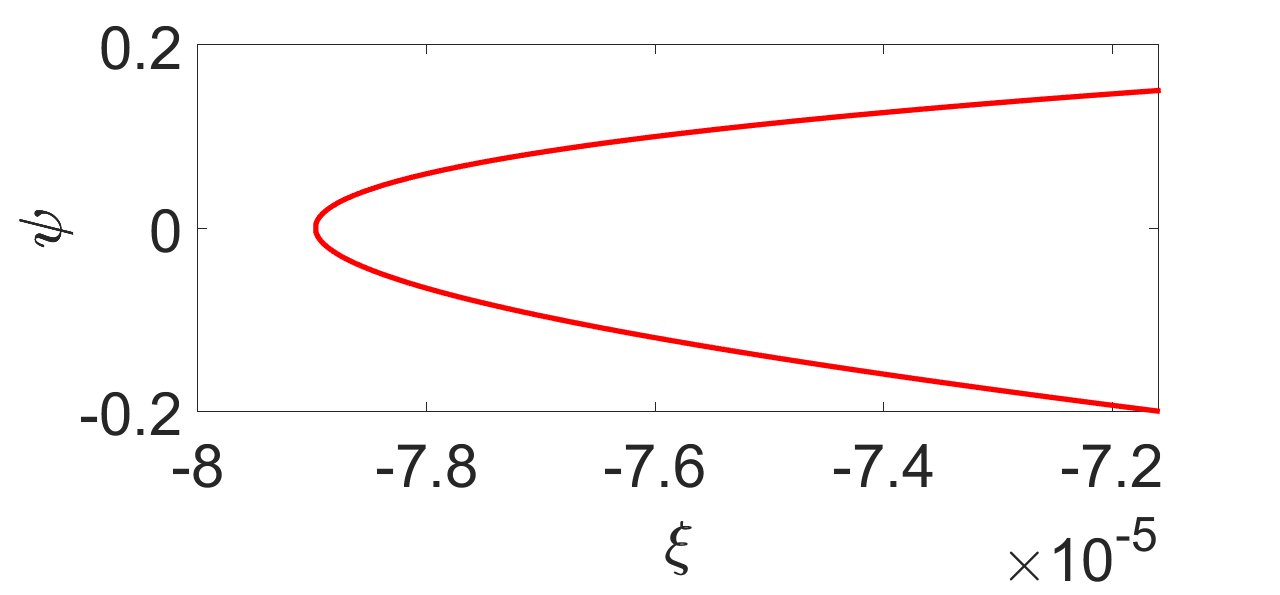}
\
\includegraphics[width=8.4cm]{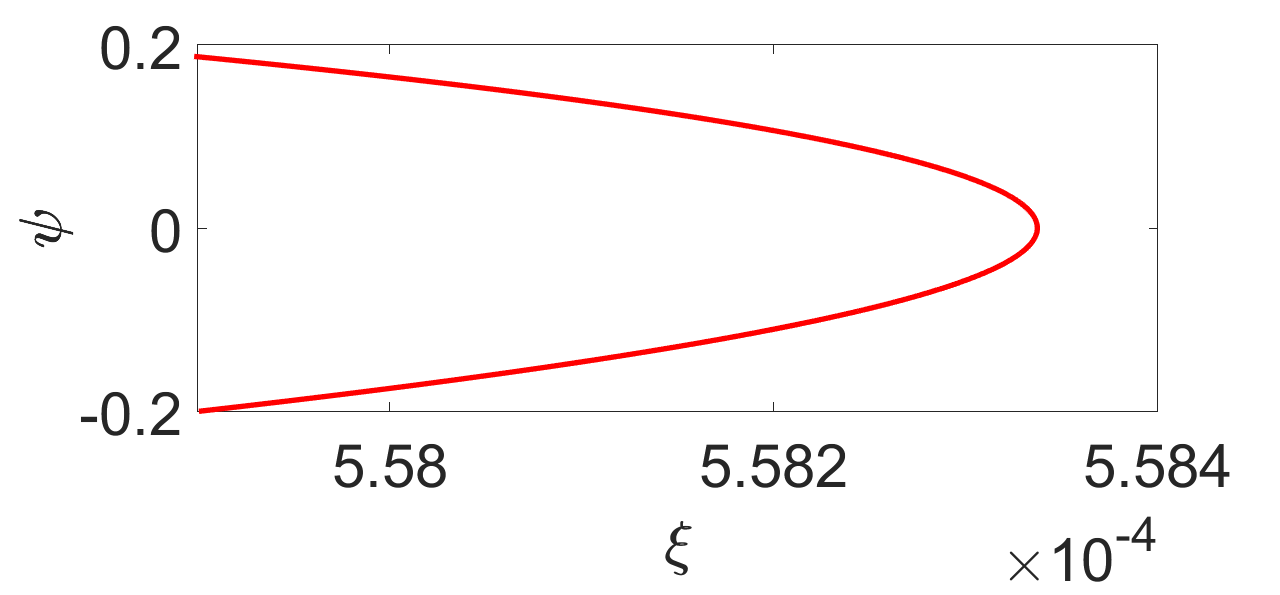}

\vspace{0.5cm}

\quad (a) $k_x=0.0001,\,k_y=0.1,L_n=0.1$ and $\omega=2$\qquad\qquad\qquad\qquad (b) $k_x=0.0001,\,k_y=1,L_n=0.1$ and $\omega=3$

\vspace{0.5cm}
\includegraphics[width=8.4cm]{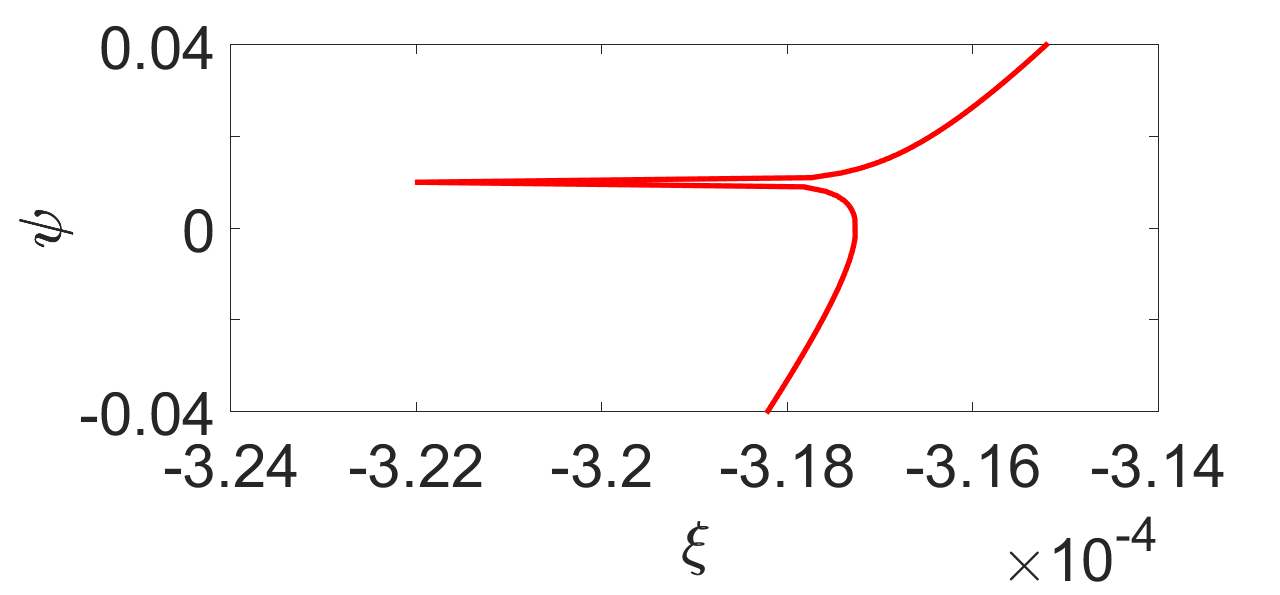}
\
\includegraphics[width=8.4cm]{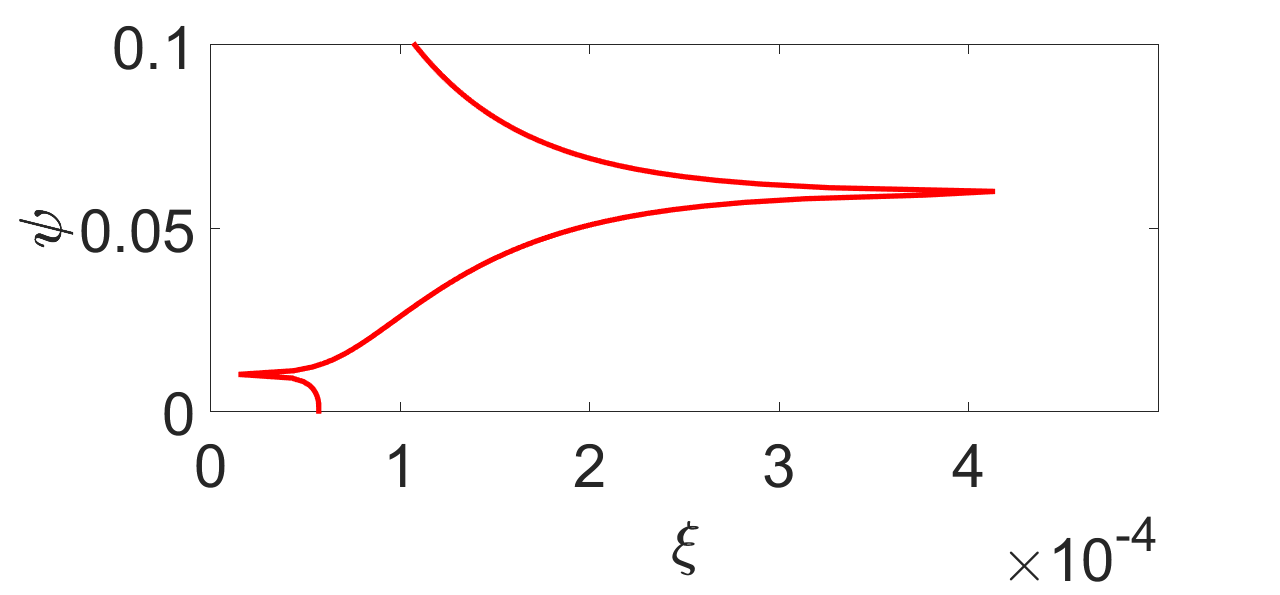}

\vspace{0.5cm}

\quad(c) $k_x=0.0001,\,k_y=2.5,L_n=0.4$ and $\omega=0.01$\qquad\qquad\qquad (d) $k_x=0.0001,\,k_y=0.1,L_n=0.8$ and $\omega=0.01$
\caption{Pictorial representation of the exact solution Eq. ({\ref{ex4_soln}}) for different values of the parameters as specified in each sub-figure. The plots in sub-figures (a)-(d) represent localized pulses for the high frequency drift waves that are similar to modulated solitary waves.}\label{fa}
\end{figure}

In the limit of very small $\psi$, i.e. $\psi<<1$, Eq. (\ref{nlc23_Finaleq}) can be written as
$$\frac{d^2\psi}{d\xi^2}+\frac{1}{\psi\left(\psi^3-\frac{k_y}{L_n}\psi^2-k^2\psi+\frac{k^2k_y}{L_n}\right)}
$$
$$
\left[\psi^3{\left(\frac{d\psi}{d\xi}\right)}^2-\frac{2k_y}{L_n}\psi^2{\left(\frac{d\psi}{d\xi}\right)}^2+k^2\psi{\left(\frac{d\psi}{d\xi}\right)}^2+\frac{k_x}{L_n}\left(\omega-\frac{k_y}{L_n}\right)\psi\frac{d\psi}{d\xi}+\psi^3-\left(\omega+\frac{2k_y}{L_n}\right)\psi^2+\frac{k_y}{L_n}\left(2\omega+\frac{k_y}{L_n}\right)\psi-\omega\frac{k_y^2}{L_n^2}\right]$$
\begin{equation}
=0. \label{ex1}
\end{equation}
As $\psi$ is very small, $\frac{1}{\psi}$ is very large. Therefore, the first term in Eq. (\ref{ex1}) is very small as compared to the rest and can be neglected. So we have
\begin{equation}
\psi^3{\left(\frac{d\psi}{d\xi}\right)}^2-\frac{2k_y}{L_n}\psi^2{\left(\frac{d\psi}{d\xi}\right)}^2+k^2\psi{\left(\frac{d\psi}{d\xi}\right)}^2+\frac{k_x}{L_n}\left(\omega-\frac{k_y}{L_n}\right)\psi\frac{d\psi}{d\xi}+\psi^3-\left(\omega+\frac{2k_y}{L_n}\right)\psi^2+\frac{k_y}{L_n}\left(2\omega+\frac{k_y}{L_n}\right)\psi-\omega\frac{k_y^2}{L_n^2}=0 \label{ex2}
\end{equation}
Again, as $\psi$ is very small, third and higher order terms in Eq. (\ref{ex2}) can be neglected to give
\begin{equation}
\frac{k_x}{L_n}\left(\omega-\frac{k_y}{L_n}\right)\psi\frac{d\psi}{d\xi}-\left(\omega+\frac{2k_y}{L_n}\right)\psi^2+\frac{k_y}{L_n}\left(2\omega+\frac{k_y}{L_n}\right)\psi-\omega\frac{k_y^2}{L_n^2}=0 \label{ex3}
\end{equation}
Solving the above first order differential Eq. (\ref{ex3}) with the method of separation of variables, we get the following exact solution:
$$
\frac{L_n\left(\omega L_n+2k_y\right)}{k_x(2\omega L_n+k_y)(\omega L_n-k_y)}\xi=\frac{1}{\sqrt{k_y(4\omega L_n-k_y)}}tan^{-1}\left[\frac{2\omega L_n+k_y}{\sqrt{k_y(4\omega L_n-k_y)}}\left(\frac{2L_n(\omega L_n+2k_y)}{k_y(2\omega L_n+k_y)}\psi-1\right)\right]+
$$
\begin{equation}
\frac{1}{2(2\omega L_n+k_y)}ln\left[{\left(\frac{2\omega L_n+k_y}{L_n}\right)}^2{\left(\frac{2L_n(\omega L_n+2k_y)}{k_y(2\omega L_n+k_y)}\psi-1\right)}^2+\frac{k_y}{L_n}\left(4\omega-\frac{k_y}{L_n}\right)\right]+C ,\label{ex4_soln}
\end{equation}
where $C$ is the constant of integration. The solution in Eq. (\ref{ex4_soln}) is similar to that derived in \cite{Acharya2026}. One interesting point that can be noticed from the exact solution Eq. (\ref{ex4_soln}) is that it is valid only form non-vanishing values of $k_x$. The nature of the exact solution Eq. (\ref{ex4_soln}) is shown in Fig. \ref{fa} for different values of parameters. Sub-figures (a), (b), (c) and (d) in Fig. \ref{fa} indicate that the high frequency electrostatic drift waves propagate in form of localized pulses for asymptotically small values of $\psi$. These typical localized pulses are pictorially similar to modulated solitary waves \cite{Yu}.

\section{Results and discussions} \label{res}
\begin{enumerate}
\item One important point in this article is that we have derived a variable coefficient second order nonlinear Eq. (\ref{nlc23_Finaleq}) to govern the dynamics of high frequency electrostatic drift waves using a travelling wave frame transformation in $2+1$ spatio-temporal dimensions without neglecting any term. This is in contrast to our earlier work \cite{Acharya,Acharya_Thesis} where we have neglected third and higher order terms to derive a nonlinear equation of third order to govern the dynamics of high frequency electrostatic drift waves in $1+1$ spatio-temporal dimensions with help of a stationary frame transformation which is subsequently decomposed into two second order equations out of which one happens to be linear and the other is nonlinear. Thus, the present article is an extension of our earlier work \cite{Acharya} from $(1+1)$ to $(2+1)$ spatio-temporal dimensions.

\item The most important point in this article is that we have performed a detailed study of the cubic dispersion relation derived using the linearized fluid equations and explicitly show various regions in the parameter space where the high frequency electrostatic drift wave becomes coupled with the ion cyclotron wave and regions where no coupling happens.  The region of growth or decay of high frequency drift waves has clearly been identified which corresponds to the coupling region of high frequency drift and ion cyclotron waves for certain intermediate values of $k_y$. Thus, the high frequency drift modes have been shown as hybrid modes generated by mixing with ion cyclotron dynamics and grow at the expense of ion cyclotron modes due to mode coupling effects. The role of the density gradient is to break the symmetry between the two ion cyclotron branches and strong values of the gradient make the branch modified by drift to overlap with the cyclotron branch leading to a mode coupling instability. 

In the case of the classic low frequency drift wave, the free energy source provided by the density gradient can lead to their destabilisation only in presence of non-adiabatic effects such as electron-ion collisions or kinetic resonance effects.  In the present work, the high frequency drift waves are not independent drift modes but are hybrid modes created by mixing with cyclotron dynamics. The linear coupling between the high frequency drift mode and ion cyclotron mode changes the system from two independent stable modes  into a coupled system where one mode  becomes unstable at the expense of the other.  It is possible that in presence of density gradient effects that store energy, as the high frequency drift wave branch approaches cyclotron frequency,  it becomes a negative energy mode and the cyclotron branch has positive energy and the coupling leads to growth of one and damping of the other.  Essentially there is a energy redistribution among the three branches.  The instability is entirely a coupling effect and happens without any Finite Larmor radius effects, or collisional effects or any other kinetic effect.

We have considered the realistic experimental values of the parameters for the excitation of the high frequency electrostatic drift wave reported in \cite{Ghosh2015,Ghosh2017} and shown the variations of the three exact roots of the cubic dispersion relation Eq. (\ref{nlc30_DR1}) corresponding to high frequency electrostatic drift and one ion cyclotron wave root in Fig. \ref{real_root}. As discussed in Sec. \ref{lin}, when the drift wave frequency becomes high viz. $\omega<\omega_{ci}$ and $\omega\geq\omega_{ci}$, for the values of $\lvert k_y \rvert$ approximately in between $0.09$ and $3$ only, the drift wave gets coupled with the ion cyclotron wave and become unstable at the expense of damping of cyclotron mode. Our results are consistent with the work reported in \cite{Ghosh2015,Ghosh2017,Klingerprl}.  
\item As discussed earlier, density gradient is the sole source of free energy driving the mode coupling instability between high frequency drift and ion cyclotron waves. In case of the classic low frequency drift waves, the free energy source provided by the density gradient can lead to their
destabilisation only in presence of non-adiabatic effects such as
electron-ion collisions or kinetic resonance effects.  In the present
work, the high frequency drift waves are not independent drift-modes but
are hybrid modes created by mixing with cyclotron dynamics. The linear
coupling between the high frequency drift mode and ion cyclotron mode
changes the system from two independent stable modes  into a coupled
system where one mode  becomes unstable at the expense of the other.  It
is possible that in presence of density gradient effects that store energy, as the high frequency drift wave branch approaches the cyclotron
frequency,  it becomes a negative energy mode and the cyclotron branch has
positive energy and the coupling leads to growth of one and damping of the
other.  Essentially there is an energy redistribution among the three
branches.  The instability is entirely a coupling effect and happens
without any Finite Larmor radius effects, or collisional effects or any
other kinetic effect.

The time-averaged energy density of a wave in a dielectric medium \cite{Landau} is
given by

$$W= \epsilon_0 \frac{{\delta E}^2}{4} \frac{d}{d\omega}[\omega
D_r(\omega)]_{\omega_r}$$

where $\delta E$ is the electric field perturbation, $D_r$ and $\omega_r$ are
real parts of the dielectric function and wave frequency.

In the present case, the real part of the dielectric function can be
written from Eq. (\ref{nlc30_DR1}) as

$$ D_r = \omega^3-(1+k_y^2)\omega+k_y/L_n = 0 $$
where we have assumed $k_x=0$.
It is interesting to focus on the values of $\omega, k_y>0$ without any
loss of generality, where a clear estimate shows that one of the branches
of cyclotron modes approximately satisfies $\omega = -\sqrt{(1+k_y^2)} $
and is quite unaffected by the density gradient.
So we can consider the following dispersion relation for the other two
modes that couple over a range of $k_y$ values
$$\omega \left (\omega-\sqrt{1+k_y^2}\right ) \left (2\sqrt{1+k_y^2}
\right ) +\frac{k_y}{L_n}= 0 $$
This can be written as

$$ \omega^2 - \omega\sqrt{1+k_y^2} +  \frac{k_y}{2 L_n\sqrt{1+k_y^2}}=0$$
The two roots of this equation are

$$\omega = \frac{\sqrt{1+k_y^2}}{2} \pm\frac{1}{2}\sqrt{(1+k_y^2)-\frac{2
k_y}{L_n}\frac{1} {\sqrt{1+k_y^2}}} $$

The energy content of the two modes depends on the sign of
$$\omega\frac{\partial D}{\partial \omega} = \omega \left
(2\omega-\sqrt{1+k_y^2} \right ) $$

It is evident that the cyclotron mode is a positive energy mode and the
high-frequency drift mode is a negative energy mode.  Therefore, the
high-frequency drift mode grows at the expense of the cyclotron mode.

\item Due to inhomogeneity in the x-direction, any behaviour in the $x$-direction can be ascertained  through non-local analysis only.  However considering very small values of $k_x$, Fig. \ref{dr3} show that the two modes nearly overlap leading to an instability for $L_n<1$. For $L_n  = 1$, the overlap region reduces to a point and then for $L_n = 1.2$, there is no overlap region and the instability completely disappears.  While for $k_x=0$, there is no instability, for $L_n\geq 1$, for small but finite values of $k_x$, the coupling of modes seems to persist upto approximately $L_n=1.1$. The threshold $L_n=1$ for the disappearance of drift cyclotron coupling is slightly modified for finite $k_x$. However, such results  can be addressed out  only through non-local analysis.

\item The values of $Ln \rightarrow Ln/\rho_s$
are usually of the order of 100 or more [5] for various tokamak devices viz. Princeton Large Torus (PLT), TEXT tokamak etc leading to the conventional drift wave frequencies with $\omega<<\omega_{ci}$. For instance, in PLT, the density gradient scale length is around $20 \,cm$ and the ion Larmor radius around $0.1\,cm$ implying the normalized $L_n\approx 200$. For the TEXT tokamak, the density gradient scale length is around $10\,cm$ implying that $L_n\approx 100$ as the ion Larmor radius is of the same order \cite{Horton_review}. However, there may also be very steep density gradients present in edge and SOL regions of tokamaks. Taking this fact into account, we consider the typical parameters of Aditya tokamak: ion sound speed $c_s=45 \, km/s$ \cite{Sangwan}; ion cyclotron frequency $\omega_{ci}=20 \,MHz$ \cite{Mishra}; un-normalized density gradient scale length $L_n=0.1\,cm$ \cite{Sahoo}. This implies that the ion sound radius is given as: $\rho_s=\frac{c_s}{\omega_{ci}}=0.225\,cm$. Therefore, we have the normalizaed density gradient scale-length, i.e. $L_n\rightarrow \frac{L_n}{\rho_s}$ is given as: $L_n=0.4$. 
It has to be emphasized here that, in \cite{Sahoo}, a range of values between $0$ and $5\,cm$ are considered for the density gradient scale-length in context of $3D$ simulations of the SOL region of Aditya tokamak. The authors \cite{Sahoo} have also considered many values of the density gradient scale length which are less than $1\,cm$ and some are approximately equal to $1\,cm$. This motivates us to take $L_n=0.1 \,cm$ in our work. As we got the normalized value of $L_n$ to be $0.4$ for the SOL region of Aditya tokamak, our results shown in Fig. \ref{dr1} become applicable. This concludes that the high frequency electrostatic drift waves can be excited in SOL regions of tokamaks that require further experimental investigations.



\item Solar corona is one of many astrophysical plasma systems where diverse values of the density gradient scale lengths can be possible \cite{Poedts2025}. We consider regions of solar corona having steeper density gradients for which the density gradient scale-length can be given as: $L_n=100\,m$ \cite{Vranjes2009a}. Also, the magnetic field strength can be as small as $B=0.1 \,G$ \cite{Poedts2025} implying the ion cyclotron frequency to be $\omega_{ci}=\frac{Be}{2\pi m_i}\approx 100\,Hz$ where $e=1.6\times10^{-19}\,C$ is the electronic charge and $m_i=1.67\times10^{-27}\,kg$ represents electronic charge and mass of Hydrogen ion respectively. In solar corona, electron temperature can be as high as $T_e=10^7\,K$ \cite{Poedts2025} which implies that the ion sound speed is given as: $c_s=\sqrt{\frac{K_BT_e}{m_i}}\approx 2.87\times10^5\,m/s$ where $K_B$ represents Boltzmann constant that is equal to $1.38\times10^{-23}\,J/K$. Thus, we have $\rho_s=2.87\times10^3\,m$ which yields the normalized density gradient scale-length to be $L_n\rightarrow\frac{L_n}{\rho_s} \approx 0.03$. For this value of $L_n$, the excitation of high frequency drift waves in solar coronal region is practically possible. From Fig. \ref{ln1}, it can be noticed that the drift wave frequency increases as the value of $L_n$ gets decreased. For $L_n=0.1$, the frequency of the drift wave reaches upto approximately $3\omega_{ci}$. If $L_n$ will be further reduced to $0.03$ as estimated considering the solar coronal conditions \cite{Vranjes2009a,Poedts2025}, the high frequency electrostatic drift wave falls in the frequency range which can even exceed $3\omega_{ci}$. Therefore, the excitations of the high frequency drift waves are feasible in realistic solar coronal conditions. 


\item
 The cubic dispersion relation for the excitation of high frequency drift waves can also be derived from Eq. (\ref{nlc23_Finaleq}). For this, Eq. (\ref{nlc23_Finaleq}) has been rewritten as:

\begin{equation}
\psi\left(\psi-\frac{k_y}{L_n}\right)\left(\psi^2-k^2\right)\frac{d^2\psi}{d\xi^2}+\psi\left(\psi^2-\frac{2k_y}{L_n}\psi+k^2\right){\left(\frac{d\psi}{d\xi}\right)}^2+\frac{k_x}{L_n}\left(\omega-\frac{k_y}{L_n}\right)\psi\frac{d\psi}{d\xi}+(\psi-\omega){\left(\psi-\frac{k_y}{L_n}\right)}^2=0. \label{mod1}
\end{equation}
Substituting the expression for $\psi$ from Eq. (\ref{nlc11}) in Eq. (\ref{mod1}) and linearizing, we get
\begin{equation}
\omega\left(\omega-\frac{k_y}{L_n}\right)\left(\omega^2-k^2\right)\frac{d^2}{d\xi^2}\left(k_xv_x+k_yv_y\right)+\frac{k_x}{L_n}\omega\left(\omega-\frac{k_y}{L_n}\right)\frac{d}{d\xi}\left(k_xv_x+k_yv_y\right)+{\left(\omega-\frac{k_y}{L_n}\right)}^2\left(k_xv_x+k_yv_y\right)=0 \label{mod2}.
\end{equation}
Now, assumptions of $v_x$ and $v_y$ to be proportional to $exp(i\xi)$ in Eq. (\ref{mod2}) leads to the following dispersion relation:
\begin{equation}
\omega^3-\left(1+k^2+i\frac{k_x}{L_n}\right)\omega+\frac{k_y}{L_n}=0, \label{mod3}
\end{equation}
which is the same dispersion relation as Eq. (\ref{nlc30_DR1}) for high frequency electrostatic drift waves. 


\item

In order to compare the normal low frequency drift and ion cyclotron wave with the three exact roots of Eq. (\ref{nlc30_DR1}), we have plotted Fig. \ref{comb1}. Similar behaviour of the high frequency drift and ion cyclotron wave roots of Eq. (\ref{nlc30_DR1}) with the normal drift and cyclotron waves can be clearly noticed from Fig. \ref{comb1}. 


\begin{figure}
\includegraphics[height=10.5cm,width=18cm]{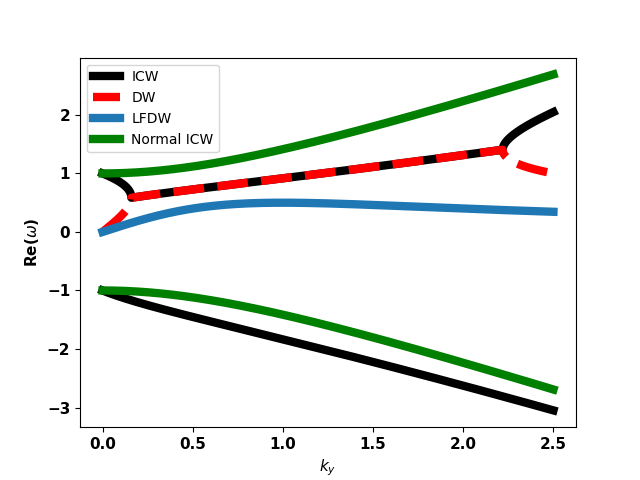}
\caption{Behaviour of the three exact roots of the dispersion relation Eq. (\ref{nlc30_DR1}) for the parameters: $k_x=0$ and $L_n=0.4$ in comparison with the conventional low frequency electrostatic drift wave for $L_n=1$ and the normal ion cyclotron wave.}\label{comb1}
\end{figure}

\item   
 Numerical solutions of the second order nonlinear Eq. (\ref{nlc23_Finaleq}) reveal periodic oscillations of the high frequency electrostatic drift waves for certain values of parameters as shown in Fig. \ref{sngplott}, Fig. \ref{sngplot} and Fig. \ref{sngplot1}. These oscillations initially show transient fluctuations of amplitudes and finally settle into periodic oscillations. In addition, for specific parameter values, the high frequency drift wave oscillations take the form of localized pulses. Exact solutions of the second order nonlinear Eq. (\ref{nlc23_Finaleq}) have been derived by decomposing it into two first order equations under certain conditions. These solutions also show the existence of localized pulses for high frequency electrostatic drift waves as shown in Fig. \ref{ad1}. This is also confirmed by the derivation of an exact analytical solution of Eq. (\ref{nlc23_Finaleq}) in the asymptotically small limit of $\psi$. This solution Eq. (\ref{ex4_soln}) pictorially indicates the presence of different localized pulses for the high frequency drift waves as shown in Fig. \ref{fa}. These localized pulses may be thought as pictorially similar to modulated solitary waves \cite{Yu}. Taking these facts into account, it can be concluded that the high frequency electrostatic drift waves propagate in the form of periodic oscillations as well as localized pulses. This is because the numerical as well as analytical solutions of Eq. (\ref{nlc23_Finaleq}) have been explored in the travelling wave frame.

\item
While a wide variety of drift-wave  modes  exist, the present study focuses on the simplest representative mode  which captures  the essential features to clearly elucidate underlying physics while facilitating its extension to the high-frequency regime.  
We have therefore considered the characteristics of electrostatic drift waves with $T_i =0$, and those propagating only  perpendicular to the magnetic field with $k_\parallel=0$ to avoid coupling with the ion-acoustic branch.   In this regime fluid models predict stable drift waves in the low-frequency regime for collisionless plasmas. In the limit when $\omega \approx \omega_{ci} $,  following kinetic analysis \cite{Ichimaru}, in the case of  negligible parallel wavenumber, a coupling between the drift wave and  ion-cyclotron wave has been shown to be effective yielding a coupled unstable drift-cyclotron mode with the instability  driven by the plasma density gradient. However, adiabatic electron response in drift waves precludes taking parallel wave number to be strictly zero.  In the long wavelength limit (small $k_\parallel$), a kinetic description becomes essential and  an interaction between resonant electrons and collisionless drift waves will lead to  destabilisation effects. The presence of ion temperature  gradients are known to  destabilise the drift waves leading  to ion temperature gradient (ITG) instability in both  fluid as well  as kinetic limits.  In the present context of the coupled drift and ion cyclotron modes exhibiting an instability,  inclusion of ion temperature gradient may be explored to study coupling with ITG instabilities.  Diamagnetic  effects arise from equilibrium  density and/or pressure gradients and play an important role in the drift wave dynamics.  The characteristic propagation frequency and direction of the drift waves are determined by the electron diamagnetic drift frequency.  However, in an ideal  fluid description, diamagnetic drifts alone are not capable of causing an instability.  The free energy source provided by the density gradient can lead to destabilisation of the drift waves only when non-adiabatic effects such as collisions, finite Larmor radius corrections  or kinetic resonances are present. When the frequency of the drift wave matches the ion-cyclotron frequency  leading to  coupled modes, the ions behave non-adiabatically, converting the energy of the diamagnetic drift motion to drive the instability.

\end{enumerate}
\section{Concluding remarks} \label{con}

In this work, high frequency electrostatic drift wave modes are found to be excited by mixing with cyclotron dynamics and grow due to mode coupling effects. The dynamical behaviour of these waves is governed by a variable coefficient second order nonlinear equation. Numerical and analytical solutions of the nonlinear equation show propagation of the high frequency drift waves in form of periodic oscillations as well as localized pulses. In some cases, the amplitudes of oscillations become modulated and also the oscillations suffer from damping as well as growing effects. Interestingly, the high frequency drift waves are found to be excited in realistic systems like tokamaks and solar corona. This has immense potential for further experimental and theoretical study.

\section{Acknowledgements}

The authors wish to acknowledge fruitful discussions with Prof. Nikhil Chakrabarti.

\end{document}